\documentclass[conference]{IEEEtran}
\IEEEoverridecommandlockouts
\usepackage{cite}
\usepackage{amsmath,amssymb,amsfonts,multirow}
\usepackage{algorithmic}
\usepackage{graphicx}
\usepackage{textcomp}
\usepackage[table,xcdraw]{xcolor}
\usepackage{ctable} 
\usepackage{subcaption}
\usepackage{xcolor}
\usepackage{soul}
\usepackage{booktabs}
\usepackage{multirow}
\usepackage{upgreek}

\def\BibTeX{{\rm B\kern-.05em{\sc i\kern-.025em b}\kern-.08em
 T\kern-.1667em\lower.7ex\hbox{E}\kern-.125emX}}
\begin{document}


\title{DISTINQT: A
Distributed Privacy Aware Learning Framework for QoS Prediction for Future Mobile and Wireless Networks}

\author{Nikolaos Koursioumpas*, Lina Magoula*, Ioannis Stavrakakis*, Nancy Alonistioti*\\ 
M. A. Gutierrez-Estevez**, Ramin Khalili**
\\
* \emph{Dept. of Informatics and Telecommunications, National and Kapodistrian University of Athens, Greece} 
\\
** \emph{Munich Research Center, Huawei Technologies Duesseldorf GmbH, Munich, Germany}
\\

\{nkoursioubas*, lina-magoula*, ioannis*, nancy*\}@di.uoa.gr \\

\{m.gutierrez.estevez**, ramin.khalili**\}@huawei.com}

\maketitle

\begin{abstract}
Beyond 5G and 6G networks are expected to support new and challenging use cases and applications that depend on a certain level of Quality of Service (QoS) to operate smoothly. Predicting the QoS in a timely manner is of high importance, especially for safety-critical applications as in the case of vehicular communications. Although until recent years the QoS prediction has been carried out by centralized Artificial Intelligence (AI) solutions, a number of privacy, computational, and operational concerns have emerged. Alternative solutions have surfaced (e.g. Split Learning, Federated Learning), distributing AI tasks of reduced complexity across nodes, while preserving the privacy of the data. However, new challenges rise when it comes to scalable distributed learning approaches, taking into account the heterogeneous nature of future wireless networks. The current work proposes DISTINQT, a novel multi-headed input privacy-aware distributed learning framework for QoS prediction. Our framework supports multiple heterogeneous nodes, in terms of data types and model architectures, by sharing computations across them. This enables the incorporation of diverse knowledge into a sole learning process that will enhance the robustness and generalization capabilities of the final QoS prediction model. DISTINQT also contributes to data privacy preservation by encoding any raw input data into highly complex, compressed, and irreversible latent representations before any transmission. Evaluation results showcase that DISTINQT achieves a statistically identical performance compared to its centralized version, while also proving the validity of the privacy preserving claims. DISTINQT manages to achieve a reduction in prediction error of up to 65\% on average against six state-of-the-art centralized baseline solutions presented in the Tele-Operated Driving use case. 
\end{abstract}


\section{Introduction}\label{Intro}

Forthcoming beyond 5G (B5G) and 6G network systems are expected to steer a new era of wireless mobile networks supporting new and challenging use cases and applications, such as autonomous vehicles, holographic communications, digital twins, pervasive intelligence and global ubiquitous connectability \cite{9349624Jiang}. Such use cases will introduce new services involving heterogeneous data streams, devices, resources, operations, and entirely new Key Performance Indicators (KPIs) \cite{9830439Moussaoui, one6G_Whitepaper_UCs}. The viability of these services is dependent on a certain level of Quality of Service (QoS) to operate smoothly and provide a satisfactory experience to end users. As such, there is a major requirement for mechanisms able to detect imminent changes in the QoS \cite{5GAA-3}, and proceed to proactive service adaptation actions, ensuring an uninterrupted service operation. The QoS may be expressed in terms of the minimum required data rate, acceptable packet loss ratio, response time, and other related KPIs.

In parallel, the use of Artificial Intelligence (AI) has been recognized as a key enabler in future wireless networks by both Industry and Academia \cite{kaloxylos_alexandros_2020_4299895}. So far, mainly centralized AI approaches \cite{Zhohov2021, Minovski2021, Sliwa2021, Gutierrez2021, 9566486Kousaridas, Palaios9605036} have been employed for QoS prediction, requiring data transfers from source devices to centrally located data centers and, thus, suffering from data privacy, computational complexity and operational costs \cite{one6G_Whitepaper_overview}. Alternative AI solutions have also recently been emerged (e.g. Split Learning, Federated Learning), distributing AI related tasks of reduced computational complexity (e.g. local training) to a number of (in general heterogeneous) devices, while exploiting locally stored datasets and avoiding any raw data exchanges \cite{NASSEF2022108820}. Overall, distributed AI solutions could contribute to effective resource utilization, scalability and privacy preservation, while reducing computational complexity and network delays.

The 3rd Generation Partnership Project’s (3GPP) in Rel. 18, attempts to adopt distributed learning by introducing multiple distributed Network Data Analytics Functions (NWDAFs) \cite{3GPP-FL}, incorporating the Model Training Logical Function (MTLF) \cite{ETSI-MTLF}. Concurrently, 3GPP’s Rel. 19, investigates new deployment and management options, as well as new functionality and KPI requirements that enable model transfer, which is of significant importance in the context of distributed learning \cite{3GPP-model_transfer}.

The benefits of distributed learning are well understood but can be gained as long as some significant challenges are addressed, related to distributing AI in wireless networks.
One of the major challenges towards an effective learning is factoring in device heterogeneity, in terms of data types, data distribution, and computational capabilities; 
model heterogeneity may also be present due to possible variations in model architectural options. The challenge in this case is to strike a good balance between the level of integration of heterogeneous model architectures and learning effectiveness.
The latter necessitates the investigation for model aggregation techniques capable of producing robust and well performing models, especially in cases of critical services and applications with stringent QoS requirements \cite{5GCroCo_D2_1}.
In addition, ways to cope with the anticipated increased burden on network resources are needed: distributed learning typically requires continuous model update transmissions, comprising millions or even billions of model parameters, a burden that increases proportionally to the number of devices participating in the distributed AI mechanism. 

In view of the limitations of centralized QoS prediction approaches and the challenges in deploying effectively distributed learning, it becomes apparent that new and innovative solutions are required for QoS prediction in future mobile networks. Such solutions should enable a scalable distributed learning execution across heterogeneous network entities and model architectures, while contributing to privacy preservation and ensuring a robust model performance.

Although distributed learning has gained significant attention, only a limited number of state-of-the-art works propose its application to predicting QoS \cite{Zhang101007,XU2023110463,Li9852729}. These works apply federated learning techniques assuming \textit{identical data types} and \textit{Neural Network (NN) architectures} across the nodes, allowing for differences only in data distribution and size. However, heterogeneous environments are not uncommon, where \textit{different types of data} may be collected by different network nodes, also requiring \textit{different NN architectures} to extract meaningful insights related to the prediction task. Such heterogeneous environments cannot be addressed by the past works and are considered in this paper.

Motivated by these open challenges, the current work proposes DISTINQT, a novel multi-headed input \cite{Harvey2015} privacy-aware distributed learning framework based on sequence-to-sequence autoencoders \cite{Sutskever2014}, capable of providing accurate predictions to end users for a future time horizon. More specifically, the DISTINQT's architecture extends upon our previous work \cite{9977782Magoula}, preserving data privacy by distributing (among distinct nodes) a number of NN-Encoders, capable of encoding input data into highly complex, compressed, and irreversible latent representations before any transmission. DISTINQT also allows different architectural options for each NN-Encoder, depending on the data types collected at the different nodes. This flexibility enables the incorporation of diverse knowledge and model architectures into a sole learning process that will enhance the robustness and generalization capabilities of the final QoS prediction model.

The key contributions can be summarized as follows:
\begin{itemize}
 \item We propose a novel distributed learning framework for QoS prediction, sharing computations of lower complexity among different nodes, training a complex multi-headed input NN architecture.
 \item Our distributed framework is able to support network entities with similar and heterogeneous characteristics, both in terms of data types and model architectures. As such, our framework captures diverse knowledge towards a robust and generalized model for QoS prediction.
 \item Our framework is privacy-aware, contributing to data privacy preservation by encoding raw input data into highly complex, compressed, and irreversible latent representations before any transmission.
 \item Our framework is evaluated in the Tele-Operated Driving use case achieving a statistically identical performance compared to its centralized counterpart and an average reduction in prediction error up to 65\% against six state-of-the-art centralized baseline solutions.
\end{itemize}

The rest of the paper is organized as follows. Section \ref{related_work} presents relevant state-of-the-art works. Section \ref{hyper_fl_framework} provides an extensive description of our proposed distributed framework that is evaluated in the Tele-operated Driving use case described in Section \ref{use_case_section}, where the performance evaluation results are provided in Section \ref{performance_evaluation}. Finally, section \ref{conclusions} concludes the paper.

\section{Related Work}\label{related_work}
This section provides an overview of the literature regarding distributed learning approaches for wireless networks, targeting also the QoS prediction. Despite the fact that the concept of distributed learning has become compelling, its application in various open challenges for future wireless networks is still at a rather early stage. 

A number of related works study different NN architectural options for distributed and federated learning. The authors of \cite{10.1145/3298981} provide a comprehensive survey on definitions, architectures, and applications for the federated-learning framework. In \cite{Errounda9842629} the authors propose a mobility vertical federated forecasting framework that allows the learning process to be jointly conducted over vertically partitioned data belonging to multiple organizations. Another work \cite{Ryu9721798} introduces the concept of split learning, reviews traditional, novel, and state-of-the-art split learning methods, and discusses current challenges and trends. The \cite{Jeon9016486} proposes a privacy-sensitive parallel split Learning method that prevents overfitting considering the differences in a training order and data size at each node. Kim \textit{et.al} \cite{Kim2021} propose a novel spatio-temporal split learning framework with multiple end-systems, in order to realize privacy-preserving Deep Neural Network (DNN) computation. The work \cite{Sun9474931} proposes an edge-enabled distributed deep learning platform by dividing a general DNN into a front and back subnetwork. The front subnetwork is deployed close to input data, trained at each edge device and all produced outputs are sent to the back subnetwork for later training at a cloud center.

There is also a number of works related to the application of distributed learning in wireless networks for a variety of use cases including computational offloading, model partitioning and energy efficiency. In works \cite{Saguil9039657} and \cite{Mohammed9155237} the authors propose a fine-grained adaptive partitioning scheme to divide a source DNN in partitions. Two integrating options are proposed including a local execution at an end device, or partition offloading to one or multiple powerful nodes. Another work \cite{Xu9234011} investigates DNN inference offloading in a 5G environment, aiming to increase the prediction request acceptance or to minimize the energy consumption of edge devices. Zeng \textit{et. al} \cite{Zeng9296560}, propose CoEdge, a distributed DNN computing system that orchestrates cooperative prediction over heterogeneous edge devices towards minimizing the system energy consumption while promising response latency requirement. In another work \cite{Du9275375} the authors propose decoupled Convolutional NN (DeCNN) structure to optimize multi-level parallelism for distributed prediction on end devices. Additionally in \cite{Hu101145}, the authors propose DeepHome, a framework that can distribute NN inference tasks to multiple heterogeneous devices inside a home. 

Finally, only a limited number of works focus on the application of distributed learning for QoS prediction in mobile networks. Zhang \textit{et.al} \cite{Zhang101007} propose a privacy-preserving QoS prediction, leveraging federated learning techniques, along with several efficiency improvements to reduce system overhead. The authors in \cite{XU2023110463} propose MultiFed, which is a multi-centre FL framework for QoS prediction via a cloud–edge collaboration mechanism. The work in \cite{Li9852729} discusses a personalized federated tensor factorization framework for distributed privacy-preserving Internet of Things (IoT) services QoS prediction. However, these works assume identical data types and NN architectures across nodes, while considering differences only in data distribution and size.

As already stated in the Introduction, there is a gap in the state-of-the-art works when it comes to distributed learning approaches for QoS prediction taking into account both common and heterogeneous nodes, in terms of data types and model architectures, along with data privacy and knowledge sharing. The current work extends \cite{9977782Magoula}, by enabling the distribution of the learning apart from the inference/prediction process. DISTINQT is the first one that proposes a distributed learning framework for QoS prediction, sharing computations of lower complexity among different nodes across the network, supporting network entities with similar and heterogeneous data types and model architectures. Our framework contributes to data privacy preservation by encoding raw input data into highly complex, compressed, and irreversible latent representations before any transmission.

\section{DISTINQT Framework} \label{hyper_fl_framework}
This section provides an extensive description of the DISTINQT framework towards a privacy preserving distributed learning scheme for QoS prediction. DISTINQT could be effectively applied to any wireless network environment comprised of distinct types of Network Entities (NETs), such as User Equipment (UE), Base Station (BS), Multi-access Edge Computing (MEC), Cloud etc. DISTINQT follows a hierarchical structure, enabling distributed learning in large-scale network environments. Following a bottom-up approach, the DISTINQT structure includes three main components: workers, aggregators and a coordinator. Each of these components has a distinct role within the structure. At the foundational level, there are multiple workers, followed by the aggregators that act as intermediaries between workers and the higher-level coordinating entity. At the top of the hierarchical structure is the coordinator, which acts as the  managing entity for the entire distributed learning system.

Let $E$ denote the number of distinct NETs. Each NET $e \in \{1,...,E\}$ may be comprised of multiple workers. All workers of a specific NET share the same input features (data types) but the number and distribution of data samples may differ. However, workers of different NETs collect and process different input features related with a learning task. 
\subsection{Role Assignment} \label{hyper_fl_framework_training}
The DISTINQT framework is realized in a synchronized manner and is capable of providing QoS predictions for a specific time horizon. At the beginning of the learning phase, each NET is assigned with one of the following roles based on data samples availability:\\
\textbf{Active NET}: An active NET holds data samples related with the learning task, as well as the set of groundtruth data (i.e. QoS KPI). Only one NET can be assigned this role.\\
\textbf{Passive NET}: A passive NET holds data samples related with the learning task, without any groundtruth data. Multiple NETs can be passive.\\
\textbf{Coordinator NET}: A coordinator NET coordinates the learning process in a synchronized manner, by contributing to the learning process with its own data samples, while also orchestrating the communication of the involved NETs. 
In our setting, the role of the coordinator is assigned to the active NET $c$, $c \in \{1,...,E\}$.\\
\textbf{Local NET Aggregator}: A local aggregator is assigned to each NET by the Coordinator NET, acting as an intermediate point in the distributed forward and backpropagation processes, by collecting and transmitting all the necessary inputs, aggregating and updating the trainable weights of the involved workers of a specific NET. 

\subsection{Neural Network Architecture and Distribution}

\begin{figure}[ht!]
 \centering
 \includegraphics[width=0.93\linewidth]{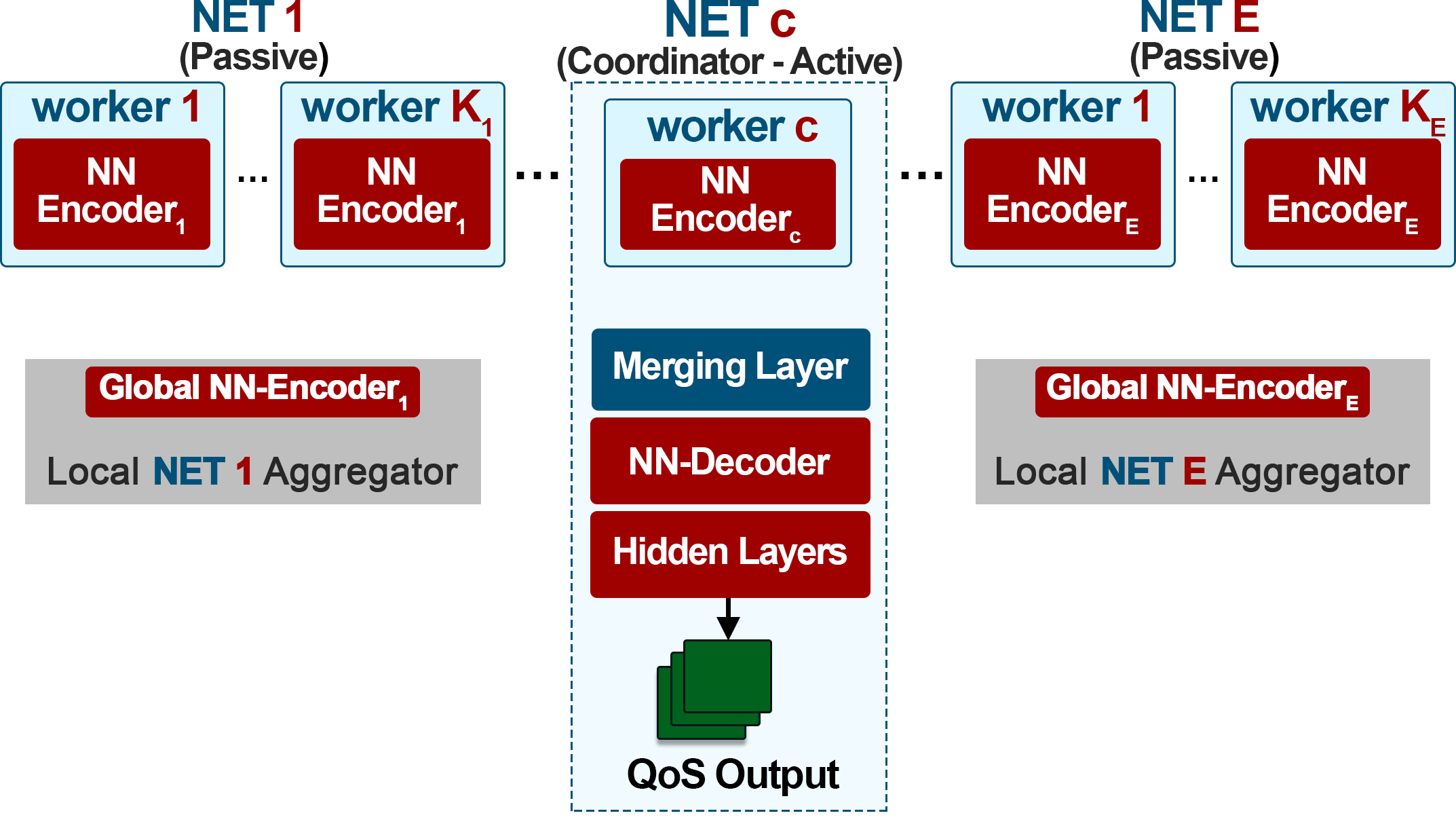}
 \caption{DISTINQT's Architecture Overview}
 \label{architecture}
 \vspace{-10pt}
\end{figure}

Fig. \ref{architecture} depicts a high level overview of the DISTINQT's NN architecture that will be distributed among all involved workers of different NETs. DISTINQT, is a novel framework that introduces a multi-headed input \cite{Harvey2015} NN architecture based on sequence-to-sequence Autoencoders \cite{Sutskever2014,8616075,10.5555/3015812.3016005,9194237}. DISTINQT is designed to encode multiple heterogeneous raw input sequences in parallel into highly complex, compressed, and irreversible latent representations (details in Section \ref{priv_section}). As a result, apart from the inherent dimensionality reduction, DISTINQT also contributes to privacy preservation, by encoding the initial raw data before any transmission. 

The structure of the DISTINQT's NN architecture (NN model layers) is as follows:\\
\textbf{NN-Encoder: } Each worker is equipped with one NN-Encoder. Let $\mathsf{K}_e$ denote the total number of workers of NET $e$, $N_{enc} = \sum_{e=1}^{E}\mathsf{K}_e$ denote the total number of NN-Encoders in the system. Each NN-Encoder may be comprised of multiple encoding layers. An NN-Encoder processes an input sequence and encodes it into a fixed-length vector, named as context vector. The workers of the same NET employ identical NN-Encoders, in terms of number and type of encoding layers, as well as number of neurons per layer. Indicative examples of such NN-Encoder types include: Long-Short Term Memory, Transformer, Convolutional encoders etc. DISTINQT shares computations of lower complexity, as the NN-Encoders used by distinct NETs may differ in their NN structure, depending on the use case, data types (e.g. visual, aural, numerical), and computational capabilities of the workers. DISTINQT's flexibility in architecture configuration could enable and support a multi-modal learning process \cite{Bayoudh2022}. 

The rest of the NN architecture is located at the coordinator NET (NET $c$) and includes a Merging Layer, an NN-Decoder, as well as a number of Hidden Layers.\\
\textbf{Merging Layer:} This layer is responsible for applying a merging function (e.g. summation, concatenation, average) to the context vectors produced by all workers, including the coordinator. The merging is performed between interconnected (associated) workers resided on different NETs (more details in Section \ref{methodology}).\\
\textbf{NN-Decoder:} The NN-Decoder is responsible for constructing the future sequence over the selected time horizon, based on the output of the Merging Layer. Contrary to its name, it should be highly noted that the NN-Decoder does not reconstruct any initial input sequence and as such cannot be considered as a privacy leak.\\
\textbf{Hidden Layers:} This part of the DISTINQT's architecture includes a set of hidden layers (e.g. fully connected), where their number and type could vary, depending on the prediction task. \\
\textbf{Global NN-Encoder:} When all NN-Encoders of a NET are updated (trainable parameters), the respective Local NET Aggregator produces a global model update, named as Global NN-Encoder, by aggregating the updated trainable parameters of the corresponding NN-Encoders. 

The Global NN-Encoder can integrate collective and heterogeneous knowledge from the participating workers, enhancing the performance, convergence speed and the generalization capabilities of our proposed framework. The latter relies on the fact that workers with limited amount of data could benefit from others with more diverse datasets. Diverse datasets frequently lead to more resilient parameter updates for the NN-Encoders that could increase the knowledge base of others with insufficient data samples in a collaborative manner (parameter aggregation).

In addition, the Global NN-Encoder also enables a seamless and effective inclusion of additional workers throughout the learning process. More specifically, if a new worker of a NET $e$ requests to participate, the responsible Local NET Aggregator will provide the latest update of the Global NN-Encoder and the new worker will join the process. In any other case, a newly added worker with a newly instantiated NN-Encoder could introduce noise in the learning process. 

\subsection{Data Privacy Preservation} \label{priv_section}

In addition to distributing a complex learning process among the involved workers of different NETs, the DISTINQT framework also aims to preserve the privacy of the initial raw data. DISTINQT allows solely the exchange of trainable parameters, gradients and context vectors between the participating NETs (Section \ref{methodology}). Similarly to FL, which was designed with privacy-awareness as a core principle, trainable parameters and gradients exchanges are not perceived as privacy violations. In the case of context vectors, a first level of de facto privacy is maintained, as they are complex latent representations of the initial raw input sequences. On top of that, each NN-Encoder of the DISTINQT further enhances the complexity of the produced context vectors by introducing non-linearity through the application of an appropriate activation function (e.g. the rectified linear activation function - ReLU) \cite{Huansheng2019,ALGULIYEV20191}. 

To further enhance the privacy of the raw data, two other methods are considered, including the application of regularization methods to the trainable parameters of each NN-Encoder and the raw input data normalization. Regularizers highly contribute to data privacy preservation by ensuring that the NN-Encoders will focus on capturing underlying data patterns, rather than memorizing specific details of the initial raw input sequences \cite{10.1007/978-3-031-25599-1_17}. Normalization indirectly contributes to data privacy preservation by not exposing the value ranges of the raw input data \cite{10.1145/3316615}. Overall, the application of all the aforementioned methods to produce the context vectors preserve the privacy of the raw input data sequences. Consequently, the produced context vectors cannot be reversed by any other worker or snooper that has gained access to the respective NN-Encoders (Section \ref{privacy_evaluation}). 

\subsection{Methodology} \label{methodology}
\begin{figure}[h!]
 \vspace{-10pt}
 \centering
 \includegraphics[width=0.85\linewidth]{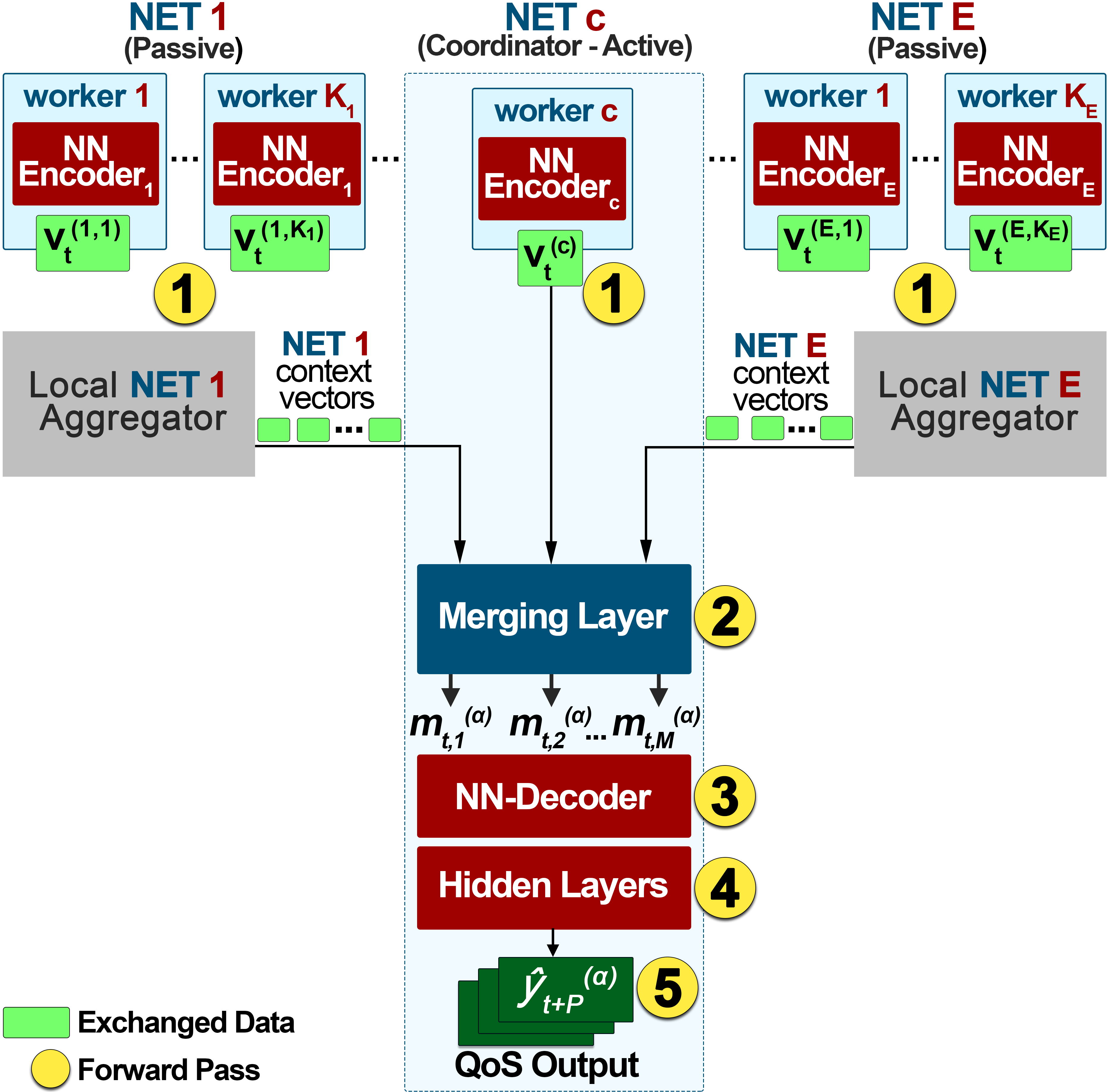}
 \caption{DISTINQT's Forward Pass}
 \label{hyper_fl_forward_pass}
 \vspace{-10pt}
\end{figure}

\begin{figure}[h!]
 \centering
 \includegraphics[width=0.85\linewidth]{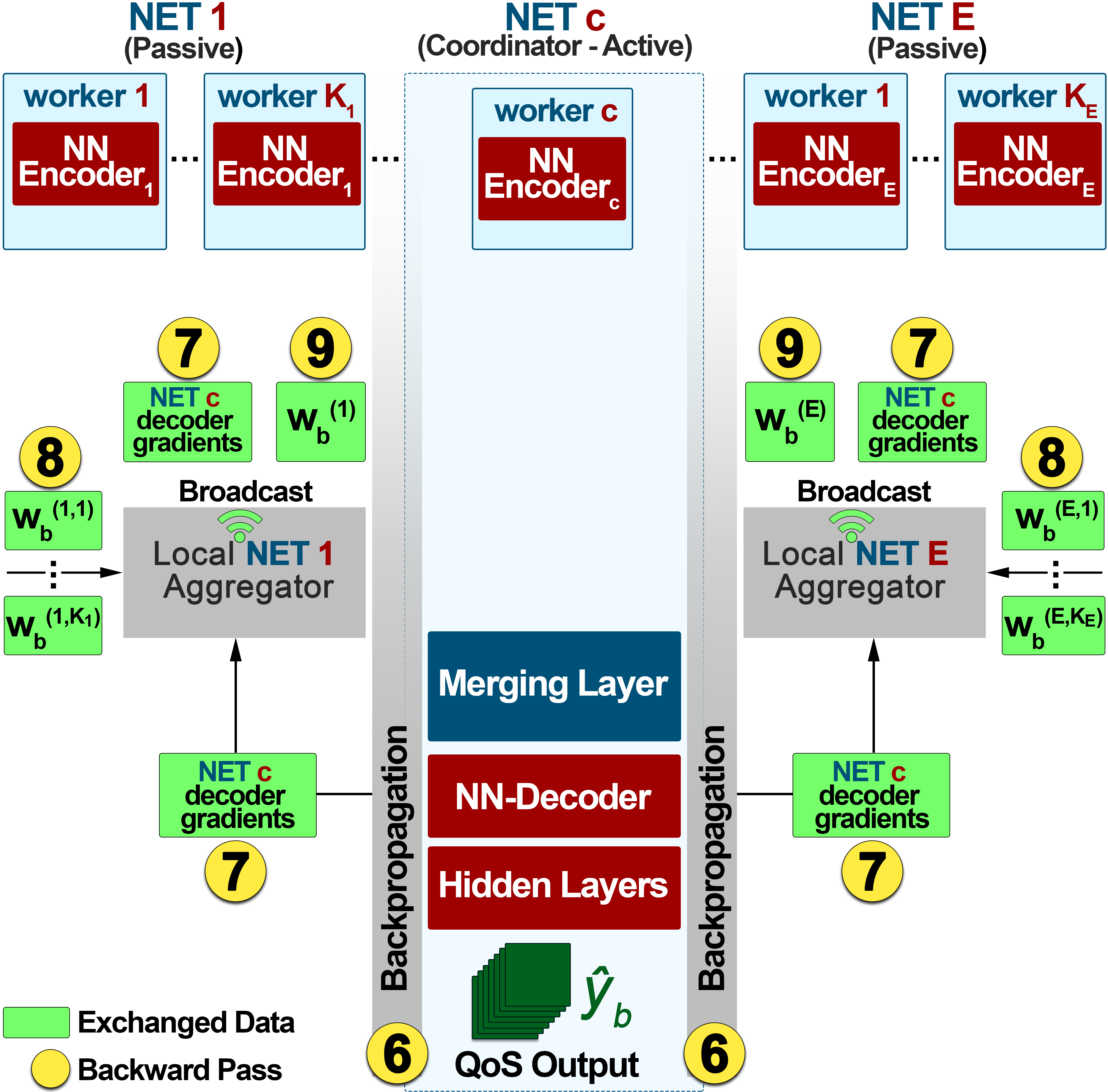}
 \caption{DISTINQT's Backward Pass}
 \label{hyper_fl_backward_pass}
 \vspace{-10pt}
\end{figure}

Figs. \ref{hyper_fl_forward_pass} and \ref{hyper_fl_backward_pass} illustrate the logical steps of the DISTINQT framework. These steps are repeated for multiple rounds (epochs) until convergence and are synchronized by the coordinator NET. Each epoch consists of multiple batch loops. In each batch loop $b$, two well-known types of processes take place, namely the forward (inference/prediction) and backward (learning) pass process. Multiple forward pass and one backward pass processes are executed during a batch loop. In each forward pass, an input feature sequence is fed through the NN in a forward direction, yielding calculations of intermediate variables and eventually generating a prediction (model output). At the end of a batch loop, a backward pass is initiated. The backward pass involves the calculations of the gradients of the loss function w.r.t the trainable parameters of each NN layer of the DISTINQT architecture by recursively applying the chain rule to update the parameters in order to minimize the loss \cite{Goodfellow-et-al-2016}. 

DISTINQT's learning procedure can be divided into nine logical steps. A complete notation table is given in Table \ref{table_notations}.\\
\textbf{Forward Pass (Fig. \ref{hyper_fl_forward_pass}):} \\
\textbf{Step \raisebox{.5pt}{\textcircled{\raisebox{-.9pt} {1}}}}: Each worker periodically collects a sequence of input features. The logging frequency of the input features $\tau_e$ ($e \in \{1,...,E\}$) may vary per NET and it is expressed through a discrete time unit, the timestep. In each batch loop $b$, all NN-Encoders process a batch $B$ of input sequences. The batch size $s$ is selected by NET $c$, based on the prediction task, and it is the same for all workers. Each input sequence initiates a forward pass process. This sequence includes historical features for $h_e$ consecutive past timesteps, comprising a historical time horizon with duration $H_e$ ($H_e = h_e\cdot\tau_e$).

\begin{figure}[h!]
 \centering
 \vspace{-10pt}
 \includegraphics[width=0.33\linewidth]{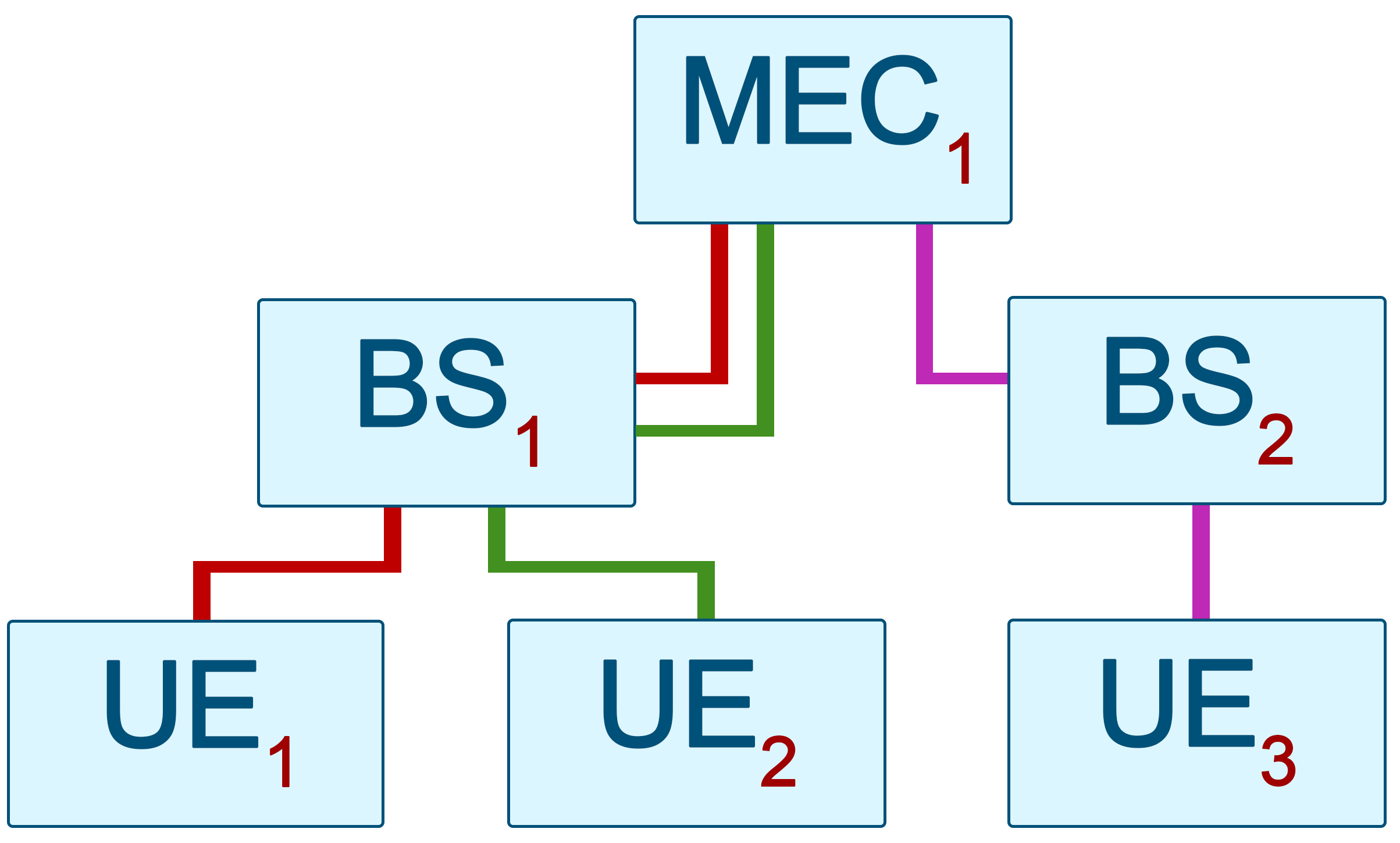}
 \caption{Example of three interconnections in a Wireless Mobile Network}
 \label{associations}
 \vspace{-20pt}
\end{figure}

\begin{figure}[h!]
 \centering
 \includegraphics[width=0.9\linewidth]{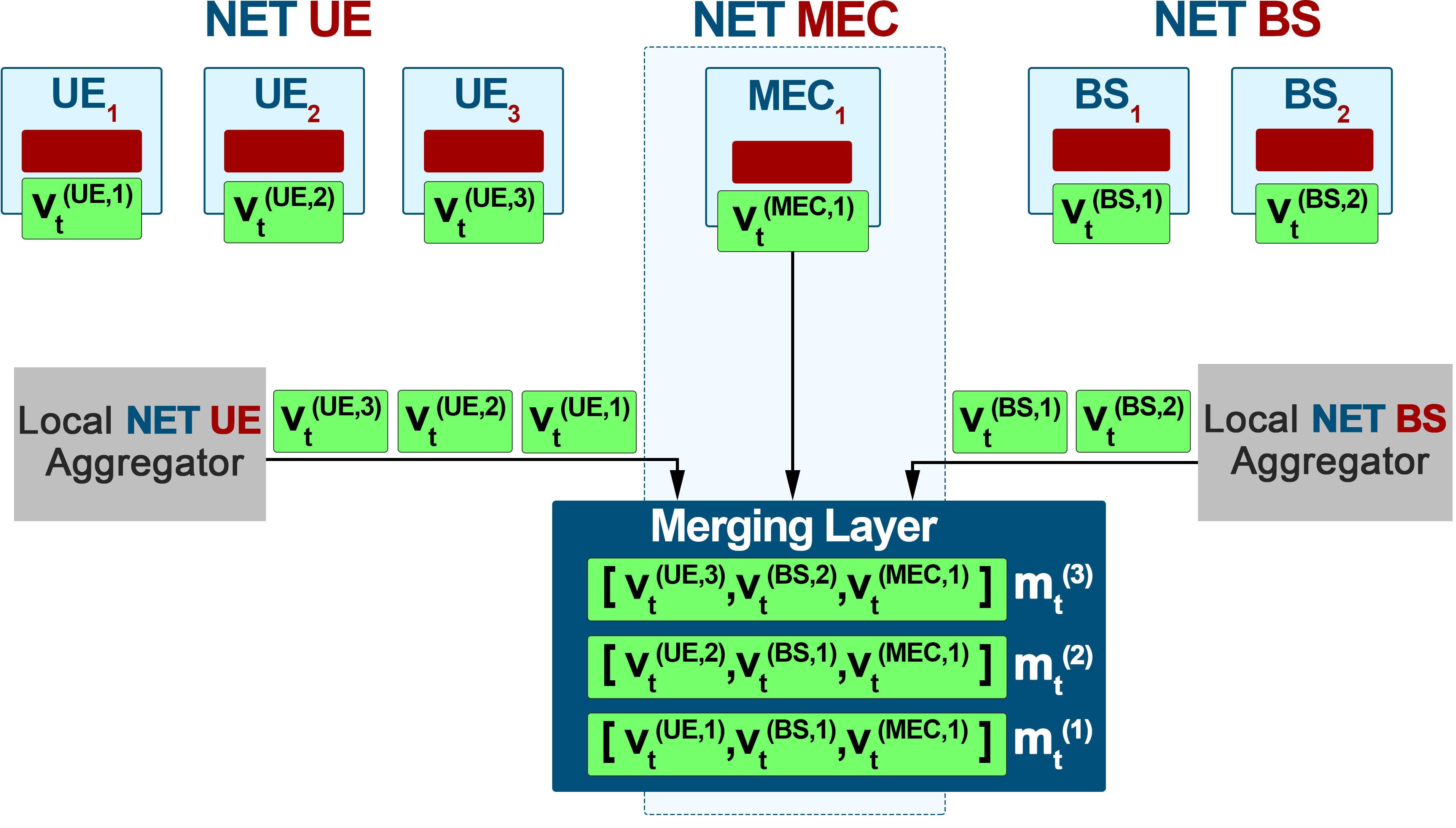}
 \caption{Example of DISTINQT's Merging Process}
 \label{merging_scheme}
 \vspace{-5pt}
\end{figure}

Let $t$ denote the timestep when all workers encode their input sequence into a context vector. We assume that the selection of $t$ is performed by the coordinator and should be greater or equal than the maximum logging frequency of all workers ($t \geq \tau_e, \forall e \in \{1,...,E\}$). This condition ensures that all workers produce context vectors only when new data samples are available, avoiding waste of resources. Let $N_e$ denote the common dimensionality (length) of the context vector of all workers of NET $e$. Let $\nu^{(e,k)}_{t,i}$, $i \in \{1,...,N_e\}$, denote the $i^{th}$ element of the context vector of worker $k$ of NET $e$ at timestep $t$. Let $\textbf{v}^{(e,k)}_{t} = [\nu^{(e,k)}_{t,1},\nu^{(e,k)}_{t,2},...,\nu^{(e,k)}_{t,N_e}]^T$ denote the context vector of worker $k$ of NET $e$ at timestep $t$. We assume that each NET worker produces a non-zero context vector at each timestep $t$ ($\textbf{v}^{(e,k)}_{t} \neq \textbf{0}$). The context vectors of all passive NETs are collected by the responsible Local NET Aggregators to transmit them to NET $c$.\\
\textbf{Step \raisebox{.5pt}{\textcircled{\raisebox{-.9pt} {2}}}}: This step includes the merging of the context vectors that is performed by the coordinator NET, which keeps track of the interconnections. The merging is realized only among the interconnected workers of different NETs. As an example, consider a wireless mobile network of three NETs, namely a UE, a BS, and a MEC, with three, two and one workers of each type, shown in Fig. \ref{associations}. The different line colors indicate the different interconnections (three in total). The MEC worker acts as the coordinator. Fig. \ref{merging_scheme} maps this wireless network to DISTINQT and illustrates the merging process. The context vector of each UE worker will be merged with the context vector of the connected (associated) BS worker and consequently with the context vector of the associated (to the respective BS) MEC-worker. As a result three merged vectors will be produced by applying a merging function to the following vectors: [$\textbf{v}_t^{(\textrm{UE},1)}$, $\textbf{v}_t^{(\textrm{BS},1)}$, $\textbf{v}_t^{(\textrm{MEC},1)}$], [$\textbf{v}_t^{(\textrm{UE},2)}$, $\textbf{v}_t^{(\textrm{BS},1)}$, $\textbf{v}_t^{(\textrm{MEC},1)}$], [$\textbf{v}_t^{(\textrm{UE},3)}$, $\textbf{v}_t^{(\textrm{BS},2)}$, $\textbf{v}_t^{(\textrm{MEC},1)}$]. 

Let $A_{t}$ denote the total number of merged vectors produced by the merging layer at timestep $t$. All merged vectors are of the same length, denoted by $M$. Let $m_{t,i}^{(a)}$ with $a \in \{1,...,A_{t}\}$ and $i \in \{1,...,M\}$ denote the $i^{th}$ element of the merged vector of interconnection $a$ at timestep $t$. Consequently, the output of the merging layer, associated with interconnection $a$, is the vector $\textbf{m}^{(a)}_t = [m_{t,1}^{(a)},...,m_{t,M}^{(a)}]^T$. In the example given by Fig. \ref{merging_scheme}, the produced merged vectors are the $\textbf{m}^{(1)}_t$, $\textbf{m}^{(2)}_t$ and $\textbf{m}^{(3)}_t$.\\
\textbf{Steps \raisebox{.5pt}{\textcircled{\raisebox{-.9pt} {3}}}}-\textbf{{\textcircled{\raisebox{-.9pt} {5}}}}: For each interconnection $a$, the merged vector $\textbf{m}^{(t)}_a$ is provided as input to the NN-Decoder. As mentioned, the NN-Decoder constructs the future sequence of the selected prediction time horizon $\mathsf{P}$ ($\mathsf{P} = p\cdot\uptau_c$) and it is comprised of $p$ prediction horizon timesteps in total. This prediction horizon timestep is selected by the coordinator and is denoted by $\uptau_c$.

\begin{table}[h!]
 \renewcommand{\arraystretch}{1.05}
 \begin{tabular}{|p{0.035\textwidth}|p{0.405\textwidth}|}
\hline
$E$ & Number of distinct NETs\\ \hline
$\mathsf{K}_e$ & Number of workers of NET $e$ \\ \hline
$B$ & Batch of input sequences \\ \hline
$b$ & Batch loop \\ \hline
$s$ & Batch size \\ \hline
$\tau_e$ & Logging frequency timestep of NET $e$\\ \hline
$\uptau_c$ & Prediction horizon timestep selected by NET $c$\\ \hline
$t$ & Timestep of the current forward pass\\ \hline
$h_e$ & Number of timesteps in a historical time horizon $H_e$\\ \hline
$p$ & Number of timesteps in a prediction time horizon $\mathsf{P}$\\ \hline
$H_e$ & Duration of historical time horizon of NET $e$\\ \hline
$\mathsf{P}$ & Duration of the prediction time horizon\\ \hline
$N_{enc}$ & Total number of Encoding layers (NN-Encoders)\\ \hline
$N_e$ & Length of the context vector of NET $e$\\ \hline
$\nu^{(e,k)}_{t,i}$ & The $i^{th}$ element of the context vector of worker $k$ of NET $e$ at timestep $t$\\ \hline
$\textbf{v}^{(e,k)}_{t}$ & Context vector of worker $k$ of NET $e$ at timestep $t$\\ \hline
$\textbf{v}^{(c)}_{t}$ & Context vector of coordinator $c$ at timestep $t$\\ \hline
$A_{t}$ & Total number of merged vectors at timestep $t$\\ \hline
$M$ & Length of the merged vector\\ \hline
$m_{t,i}^{(a)}$ & The $i^{th}$ element of the merged vector of interconnection $a$ at timestep $t$\\ \hline
$\textbf{m}^{(a)}_t$ & Vector output of the merging layer\\ \hline
$\hat{\textbf{y}}^{(a)}_t$ & Predicted output sequence of interconnection $a$ at timestep $t$\\ \hline
$\hat{\textbf{y}}_{b}$ & Set of predicted output sequences at batch loop $b$\\ \hline
$\textbf{w}_b^{(e,k)}$ & The weight updates of worker $k$ of NET $e$ at batch loop $b$ \\ \hline
$\textbf{w}_b^{(e)}$ & Aggregated trainable weights of NET $e$ at batch loop $b$ \\ \hline
\end{tabular}
\caption{Notation Table} 
\label{table_notations} 
\vspace{-10pt}
\end{table}

The output of the NN-Decoder feeds the next Hidden Layers. The final output is a sequence of the selected QoS KPI over the $p$ future timesteps, denoted as $\hat{\textbf{y}}^{(a)}_t = [\hat{y}^{(a)}_{(t+\tau)},...,\hat{y}^{(a)}_{(t+p\cdot \uptau_c)}]^T$. The set of outputs produced at batch loop $b$ is defined as $\hat{\textbf{y}}_{b} = \{\hat{\textbf{y}}^{(a)}_t, \forall t \in B, \forall a\}$ and its size is equal to $s \times a$.\\
\textbf{Backward Pass (Fig. \ref{hyper_fl_backward_pass}):} 
\\\textbf{Step \raisebox{.5pt}{\textcircled{\raisebox{-.9pt} {6}}}}: Once the forward pass processes of batch loop $b$ are complete, the predicted output sequence $\hat{\textbf{y}}_{b}$ along with the groundtruth sequence will be used, in order to minimize the loss function of the DISTINQT framework and perform the backpropagation process. During the backpropagation, the gradients of each layer are computed by recursively applying the chain rule up to the NN-Encoder of worker $c$. Gradients are the partial derivatives of the loss function, indicating the rate of change of the loss function with respect to each trainable parameter. These computed gradients will be used in order to update the trainable weights of worker $c$.\\
\textbf{Step \raisebox{.5pt}{\textcircled{\raisebox{-.9pt} {7}}}}: The computed gradients of the NN-decoder are forwarded to the Local NET Aggregators of the involved NETs to continue a distributed backpropagation process. The Local NET Aggregators broadcast the received gradients to the workers of the associated NET to update their trainable weights using their own optimizers. Let $\textbf{w}_b^{(e,k)}$ denote the weight updates of worker $k$ of NET $e$ at the end of batch loop $b$. \\
\textbf{Step \raisebox{.5pt}{\textcircled{\raisebox{-.9pt} {8}}}}: All workers send their updated weights to the respective Local NET Aggregator for aggregation based on an averaging function. The Local NET Aggregator of each NET $e$ updates the Global NN-Encoder with the aggregated weights $\textbf{w}_b^{(e)}$, given by: \\ 
\vspace{-10pt}
\begin{equation} \label{weight_aggregation}
 \textbf{w}_b^{(e)} = \frac{\sum_{k=1}^{\mathsf{K}_e}\textbf{w}_b^{(e,k)}}{\mathsf{K}_e}.
 \vspace{-8pt}
\end{equation}\\
\textbf{Step \raisebox{.5pt}{\textcircled{\raisebox{-.9pt} {9}}}}: The Local NET Aggregator broadcasts the aggregated trainable weights $\textbf{w}_b^{(e)}$ back to the respective workers to update their NN-Encoder. This step completes a batch loop of the learning process. It should be noted that after this step, all workers of a NET will have the same NN-Encoder, i.e. same trainable weights.

Steps \textbf{\raisebox{.5pt}{\textcircled{\raisebox{-.9pt} {1}}}-\raisebox{.5pt}{\textcircled{\raisebox{-.9pt} {9}}}} are repeated for each batch loop and for each epoch of the learning process until convergence. Finally, it should be mentioned that since the learning process is repeated for each batch loop, it could potentially lead to an increase in the communication overhead. This overhead could be potentially mitigated by applying effective compression techniques for NNs, such as connection pruning, weight quantization, knowledge distillation \cite{MARINO2023152}. However, since this aspect is not covered in this work, we assume that there are no strict limitations in communication resources.

\section{DISTINQT in Tele-Operated Driving} \label{use_case_section}
The current section presents the Tele-Operated Driving (ToD), \cite{5GAA-1}, and simulation environments used in the evaluation of the DISTINQT framework, along with the considered configuration setups. The proposed DISTINQT framework will be applied in order to predict the uplink throughput of a ToD-UE over a future time horizon; this time horizon is set equal to 20 secs for the derivation of the results presented in the next sections. 

\subsection{Configuration Setup}

\textbf{Network Environment:}
The discrete event network simulator (NS3) \cite{ns3} has been used for the performance evaluation of the DISTINQT framework. The simulated topologies and the evaluation methodology chosen for this study are in line with the 3GPP's guidelines \cite{3GPP_TR_38_913}, \cite{3GPP_TR_38_885}. More specifically, the simulated topology chosen for this study is an Urban Macro deployment for connected cars, known as \textit{Manhattan Grid}. The duration of the simulation is equal to 280 minutes, while the logging frequency of the data is equal to $\tau_e=200$ milliseconds (ms) for all NETs in our study. Two types of UEs are considered in the simulation, namely ToD-UE and NonToD-UE vehicles, one BS and one MEC server. According to \cite{5GAA_ToD}, the periodic traffic of the ToD-UEs is set to 20 Mbps for the uplink and 500 Kbps for the downlink. On the other hand, the NonToD-UEs generate periodic background traffic load that interferes with that from the ToD-UE vehicles and will impact on the latter's QoS. The number of NonToD-UEs varies over time and is randomly selected in the range of $[0, 100]$. Each NonToD-UE generates traffic of 4 Mbps for the uplink and 8 Kbps for the downlink, considered to be constant throughout the experiments. The mobility of all UEs was generated using the Simulation of Urban Mobility tool (SUMO) \cite{sumo}. The ToD-UEs are moving continuously with a variable speed in the range of $(0, 14]$ m/s corresponding to $(0, 50]$ km/h, while the NonToD-UEs follow randomly generated trajectories inside the simulated topology, as provided by SUMO. We consider 10 simulated experiments, differing in the total background traffic load (depending on the number of NonToD-UEs) and the mobility patterns of the UEs. Table \ref{ns3Parameters} summarizes the NS-3 parameters used in the simulated scenario.
\vspace{-5pt}
\begin{table}[h!]
 \centering
\renewcommand{\arraystretch}{1}
 \begin{tabular}{c c}
 \hline
 \textbf{Parameter Description} & \textbf{Default Value} \\ [0.7ex]
 \hline %
 No. of ToD-UES \& BSs \& MECs & 5 \& 1 \& 1\\
 \hline %
 No. of NonToD-UES & [0, 100]\\
 \hline %
 UE \& BS Transmission Power & 23 dBm \& 40 dBm\\
 \hline %
 BS Center Frequency & 2160 MHz \\
 \hline %
 BS Downlink and Uplink Bandwidth & 20 MHz \\
 \hline %
 BS and UE antenna model type & 1x Isotropic \\
 \hline %
 Propagation Loss Model for LOS/NLOS & ITU-R P.1411 \\
 \hline %
 \end{tabular}
 \caption{Network Simulation Parameters}
 \label{ns3Parameters}
 \vspace{-9pt}
\end{table}

\textbf{Input Feature Configuration}:
Three distinct NETs are considered in this work, namely the ToD-UE, the BS and the MEC Server. The ToD-UE NET is comprised of 5 workers (one for each of the 5 ToD-UEs), while the BS and the MEC server have 1 worker each. The BS is an active NET (i.e., holds the groundtruth) and also the Coordinator NET (i.e., coordinating the learning process). The rest of the NETs are passive. The set of input features for QoS prediction collected at each NET, are listed below: \\
 \textbf{ToD-UE NET}:
 \begin{itemize}
 \itemsep0em 
 \item location of the ToD-UE (\textit{latitude}, \textit{longitude}) 
 \item speed of the ToD-UE in $m/s$
 \item distance of the ToD-UE from BS in $m$ 
 \item future location of the ToD-UE at the end of the selected prediction time horizon (\textit{latitude}, \textit{longitude})
 \end{itemize}
 \textbf{BS NET}:
 \begin{itemize} 
 \itemsep0em 
 \item BS traffic load, in terms of used Resource Blocks
 \item number of vehicles connected to the BS, in which the ToD-UE is attached to
 \item uplink SINR of the ToD-UE in $dBm$ 
 \item uplink throughput of the ToD-UE in $Mbps$
 \end{itemize}
 \textbf{MEC Server NET}:
 \begin{itemize} 
 \itemsep0em 
 \item number of vehicles attached to neighboring BSs
 \item traffic load, in terms of used Resource Blocks, of neighboring BSs
 \end{itemize}

We consider two distinct configuration setups, as listed in Table \ref{features_configuration}. Configuration $\textbf{c1}$ includes solely the input features from the ToD-UE and the BS. Configuration $\textbf{c2}$ includes the input features of all considered NETs (i.e. also the MEC server NET). These configurations are specifically selected to demonstrate the applicability and the model performance of our DISTINQT framework, as a higher number of (heterogeneous) NETs are collaboratively involved in the learning process.
\vspace{-7pt}
\begin{table}[h!]
 \centering
 \begin{tabular}{c c}
 \hline
 \textbf{Configuration} & \textbf{Input Features} \\ [0.7ex]
 \hline %
 \textbf{c1} & ToD-UE $+$ BS \\
 \hline %
 \textbf{c2} & ToD-UE $+$ BS $+$ MEC Server\\
 \hline %
 \end{tabular}
 \caption{Input Feature Configuration}
 \label{features_configuration}
 \vspace{-5pt}
\end{table}

\textbf{DISTINQT's Architecture Configuration: } For the selected ToD use case, the architecture of the DISTINQT framework incorporates a Bidirectional Long Short Term Memory (BiLSTM) Layer as an NN-Encoder, a Concatenation Layer for merging, an LSTM NN-Decoder, followed by a Fully Connected (FC) Hidden Layer. For each configuration, the DISTINQT’s related parameters were optimized by exploiting a custom grid search method, with over 200 combinations of different model parameters. The selected parameters include: number of neurons per layer, L2 (weight decay) regularizer \cite{Goodfellow-et-al-2016}, activation function (AF) per layer, type of optimizer, learning rate (\textit{lr}) and batch size.
With respect to the evaluation metrics explained and provided in Section \ref{performance_evaluation}, the best scored parameter combination of the DISTINQT's architecture for $\textbf{c1}$ and $\textbf{c2}$ configurations is shown in Table \ref{tab:grid_search}. The best scored parameter combination yielded NN-Encoders with the same number of neurons for all workers. To keep the presentation simple, the "Neurons" parameter shown in Table \ref{tab:grid_search} has the following form: \{NN-Encoder\} \textbf{x} \{NN-Decoder\} \textbf{x} \{FC\}.

\begin{table}[h!]
\centering
\begin{tabular}{|c|cc|}
\hline
\cellcolor[HTML]{EFEFEF}{\textbf{Parameter}} & \multicolumn{1}{c|}{\cellcolor[HTML]{EFEFEF}{\textbf{c1}}} & \cellcolor[HTML]{EFEFEF}{\textbf{c2}} \\ \hline
\textbf{Neurons}  & \multicolumn{1}{c|}{100\textbf{x}300\textbf{x}200} & 100\textbf{x}200\textbf{x}100 \\ \hline
\textbf{L2}  & \multicolumn{1}{c|}{1e-06}  & 1e-08  \\ \hline
\textbf{AF (All Layers)} & \multicolumn{2}{c|}{ReLU}  \\ \hline
\textbf{Optimizer \& lr} & \multicolumn{1}{c|}{Adam \& 1e-03} & Adam \& 5e-05 \\ \hline
\textbf{Batch Size $s$} & \multicolumn{2}{c|}{64}  \\ \hline
\textbf{Encoder Complexity (MFLOPs)} & \multicolumn{1}{c|}{90.5} & 90.5\\ \hline
\textbf{Total Complexity (MFLOPs)} & \multicolumn{1}{c|}{906} &  816 \\ \hline
\end{tabular}%
\caption{ DISTINQT's Configuration Parameters: \textbf{c1} \& \textbf{c2}}
\label{tab:grid_search}
\vspace{-5pt}
\end{table}
To showcase the distribution of lower complexity computations, Table \ref{tab:grid_search} also includes the complexity of the total DISTINQT architecture (\textit{Total Complexity}) along with the complexity of each NN-Encoder (\textit{Encoder Complexity}) in MFLOPs (Mega Floating Point Operations)\cite{torch_utils}. As it can be deduced from the table, the complexity of each NN-Encoder is approximately 10 times lower than that of the entire architecture.  
The maximum number of epochs for the learning phase was set to 1000. A min-max scaling method was applied to rescale the initial raw input data in the range $[0,1]$. Additionally, in order to avoid overfitting and potential degradation in the performance of the model, an early stopping function is applied, which monitors the Mean Squared Error (MSE) of the validation set \cite{Prechelt1998}. As a result, the learning process stops, when the MSE score does not improve or starts deteriorating. 

\section{Performance Evaluation} \label{performance_evaluation}
 In the selected ToD use case, the DISTINQT framework is applied in order to predict the uplink throughput of a ToD-UE over a prediction (future) time horizon of $\mathsf{P}=20$ secs. 
 The historical time horizon of monitored features, which are inputs to DISTINQT, is set to $H_e = 25$ secs for all NETs; after a thorough experimentation, this value was determined to yield the best performance. For the learning phase, each NET uses 3280 seconds of input data. In the evaluation study, the prediction process moves forward in time in steps of size $t$ secs. That is, every $t=1$ sec the framework provides the predicted uplink throughput for the next prediction horizon.
The proposed DISTINQT was evaluated by employing the Mean Absolute Error (MAE) metric (in Mbps), supplemented by its standard deviation, \cite{mae}, based on 10 different evaluation scenarios (ES) per ToD-UE (50 ESs in total) under diverse background traffic load levels, with a mean simulation duration equal to $\sim$1116 seconds. These scenarios were selected to assess the robustness of our framework, under different network traffic conditions. The MAE score is calculated for each prediction horizon timestep ($\tau_c=200$ ms in our study for all types of NETs) and for all prediction time windows. Each prediction time window consists of $\frac{\mathsf{P}\text{(ms)}}{\tau_c} = p$ prediction timesteps, where $p =\frac{20000}{200}= 100$; these prediction timesteps associated with the current timestep, $t$, are given by the sequence ($\{t+200, t+400,..., t+20000\}$ in ms; the historical time window $H_e$ secs corresponds to $h_e$ = 125 timesteps of historical information.
 

\subsection{DISTINQT vs Centralized}

Although centralized learning solutions present practical limitations (Section \ref{Intro}), they are considered as high performing architectures in a variety of problems and use cases. As such, we compare the performance of our DISTINQT framework with a centralized LSTM-based approach, introduced in our previous work \cite{BarMagKour2021}. Table \ref{tab:centr_vs_distr_all_scenarios}, presents the evaluation results as average values of the MAE score for both input feature configurations and for all evaluation scenarios (50 in total). Two types of evaluation metrics are considered: (a) The "overall MAE" score, representing the average MAE score over all prediction horizon timesteps, and (b) the "20th Second MAE" score, representing the average MAE score of the last prediction horizon timestep only. The results under both the centralized, \cite{BarMagKour2021}, and our proposed DISTINQT approaches are statistically very similar, with the DISTINQT showcasing a marginally better performance of 7.75\% on average for both configurations and types of scores considered. Also \textbf{c2} improves upon \textbf{c1} on average by 5.65\% for both scores, highlighting the importance of the additional input features of the MEC Server NET. MAE scores capture the overall ability of the model to yield a relatively low prediction error under the DISTINQT and the centralized approaches, considering typical uplink throughput values of the ToD applications (10-50 Mbps), as indicated by the EU funded 5GCroCo project \cite{5GCroCo_D2_1}.

\begin{table}[h!]
\centering
\resizebox{160pt}{!}{%
\begin{tabular}{|
>{\columncolor[HTML]{EFEFEF}}l |cl|cl|}
\hline
\multicolumn{1}{|c|}{\cellcolor[HTML]{EFEFEF}\textbf{\begin{tabular}[c]{@{}c@{}}Avg. (Mbps)\\ ($\pm$ STD)\end{tabular}}} &
 \multicolumn{2}{c|}{\cellcolor[HTML]{EFEFEF}\textbf{\begin{tabular}[c]{@{}c@{}}Overall\\ MAE\end{tabular}}} &
 \multicolumn{2}{c|}{\cellcolor[HTML]{EFEFEF}\textbf{\begin{tabular}[c]{@{}c@{}}20th Second\\ MAE\end{tabular}}} \\ \hline
\cellcolor[HTML]{EFEFEF}\textbf{DISTINQT (c1)} &
 \multicolumn{2}{c|}{\cellcolor[HTML]{FFFFFF}\begin{tabular}[c]{@{}c@{}}0.554\\ ($\pm$1.578)\end{tabular}} &
 \multicolumn{2}{c|}{\cellcolor[HTML]{FFFFFF}\begin{tabular}[c]{@{}c@{}}0.6854\\ ($\pm$1.4340)\end{tabular}} \\ \hline
\textbf{Centralized (c1)} &
 \multicolumn{2}{c|}{\cellcolor[HTML]{FFFFFF}\begin{tabular}[c]{@{}c@{}}0.6358\\ ($\pm$1.795)\end{tabular}} &
 \multicolumn{2}{c|}{\cellcolor[HTML]{FFFFFF}\begin{tabular}[c]{@{}c@{}}0.734\\ ($\pm$1.389)\end{tabular}} \\ \hline
\cellcolor[HTML]{EFEFEF}\textbf{DISTINQT (c2)} &
 \multicolumn{2}{c|}{\cellcolor[HTML]{FFFFFF}\begin{tabular}[c]{@{}c@{}}0.529\\ ($\pm$1.574)\end{tabular}} &
 \multicolumn{2}{c|}{\cellcolor[HTML]{FFFFFF}\begin{tabular}[c]{@{}c@{}}0.6388\\ ($\pm$1.325)\end{tabular}} \\ \hline
\textbf{Centralized (c2)} &
 \multicolumn{2}{c|}{\cellcolor[HTML]{FFFFFF}\begin{tabular}[c]{@{}c@{}}0.568\\ ($\pm$1.667)\end{tabular}} &
 \multicolumn{2}{c|}{\cellcolor[HTML]{FFFFFF}\begin{tabular}[c]{@{}c@{}}0.6702\\ ($\pm$1.303)\end{tabular}} \\ \hline
\end{tabular}%
}
\caption{DISTINQT vs Centralized QoS prediction performance for input feature configurations \textbf{c1} \& \textbf{c2}}
\label{tab:centr_vs_distr_all_scenarios}
\vspace{-10pt}
\end{table}

\begin{figure*}[h!]
 \centering
 \includegraphics[width=0.9\linewidth]{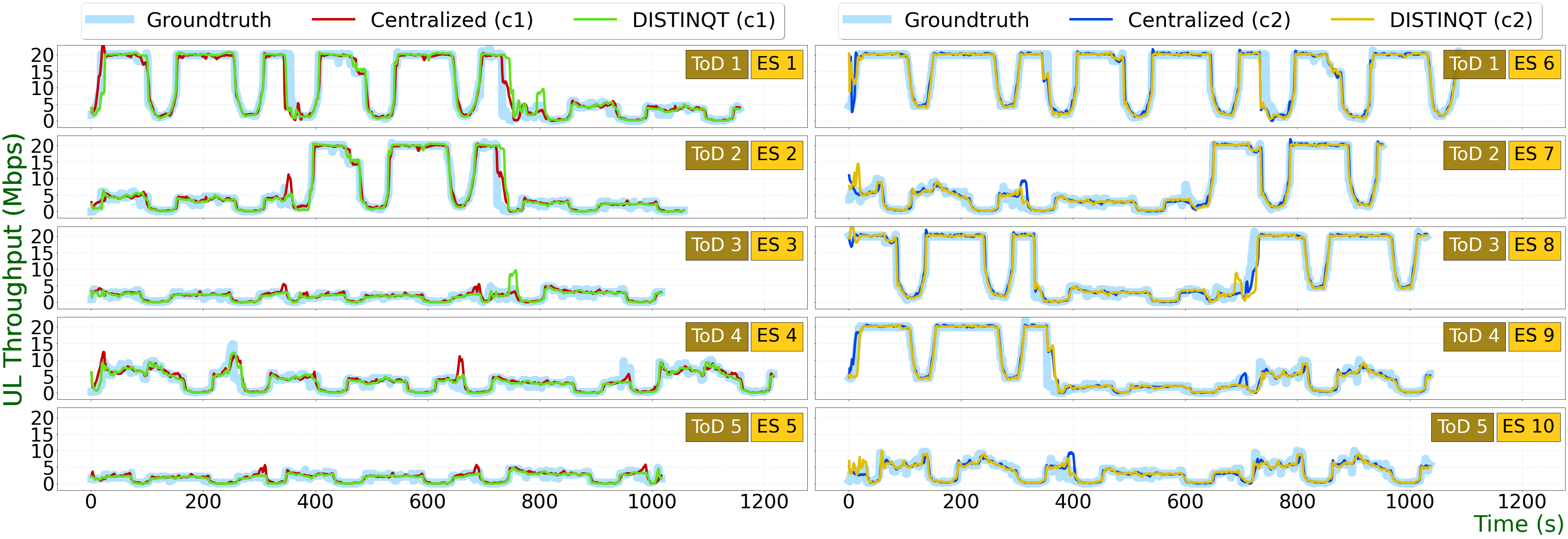}
 \caption{Groundtruth vs Predicted (Centralized and DISTINQT) uplink throughput in the 20th second of the prediction horizon}
 \label{act_pred_performance}
\end{figure*}

\begin{figure}[h!]
 \centering
 \includegraphics[width=0.9\linewidth]{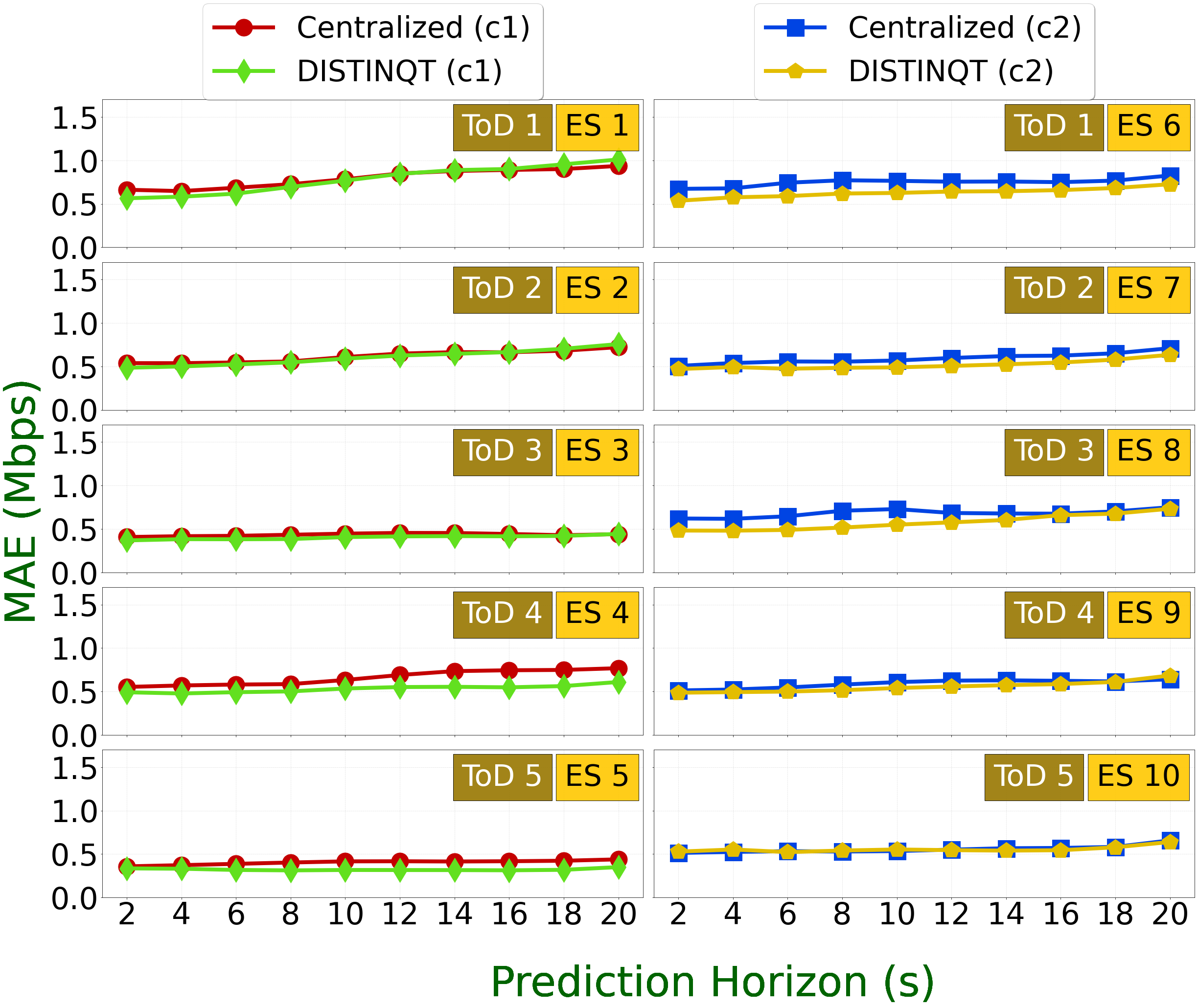}
 \caption{QoS prediction performance of DISTINQT and Centralized schemes over different prediction time horizons}
 \label{distr_centr_c1_vs_c2}
 \vspace{-15pt}
\end{figure}

\begin{table*}[h!]
\centering
\renewcommand{\arraystretch}{1.05}
\resizebox{480pt}{!}{%
\begin{tabular}{|l|cc|cc|l|cc|cc|}
\hline
\rowcolor[HTML]{EFEFEF} 
\multicolumn{1}{|c|}{\cellcolor[HTML]{EFEFEF}} &
 \multicolumn{2}{c|}{\cellcolor[HTML]{EFEFEF}\textbf{DISTINQT (c1)}} &
 \multicolumn{2}{c|}{\cellcolor[HTML]{EFEFEF}\textbf{Centralized (c1)}} &
 \multicolumn{1}{c|}{\cellcolor[HTML]{EFEFEF}} &
 \multicolumn{2}{c|}{\cellcolor[HTML]{EFEFEF}\textbf{DISTINQT (c2)}} &
 \multicolumn{2}{c|}{\cellcolor[HTML]{EFEFEF}\textbf{Centralized (c2)}} \\ \cline{2-5} \cline{7-10} 
\multicolumn{1}{|c|}{\multirow{-2}{*}{\cellcolor[HTML]{EFEFEF}\textbf{\begin{tabular}[c]{@{}c@{}}Avg. (Mbps)\\ ($\pm$ STD)\end{tabular}}}} &
 \multicolumn{1}{c|}{\textbf{\begin{tabular}[c]{@{}c@{}}Overall\\ MAE\end{tabular}}} &
 \textbf{\begin{tabular}[c]{@{}c@{}}20th Second \\ MAE\end{tabular}} &
 \multicolumn{1}{c|}{\textbf{\begin{tabular}[c]{@{}c@{}}Overall\\ MAE\end{tabular}}} &
 \textbf{\begin{tabular}[c]{@{}c@{}}20th Second \\ MAE\end{tabular}} &
 \multicolumn{1}{c|}{\multirow{-2}{*}{\cellcolor[HTML]{EFEFEF}\textbf{\begin{tabular}[c]{@{}c@{}}Avg. (Mbps)\\ ($\pm$ STD)\end{tabular}}}} &
 \multicolumn{1}{c|}{\textbf{\begin{tabular}[c]{@{}c@{}}Overall\\ MAE\end{tabular}}} &
 \textbf{\begin{tabular}[c]{@{}c@{}}20th Second \\ MAE\end{tabular}} &
 \multicolumn{1}{c|}{\textbf{\begin{tabular}[c]{@{}c@{}}Overall\\ MAE\end{tabular}}} &
 \textbf{\begin{tabular}[c]{@{}c@{}}20th Second \\ MAE\end{tabular}} \\ \hline
\rowcolor[HTML]{FFFFFF} 
\cellcolor[HTML]{EFEFEF}\textbf{ToD 1, ES 1} &
 \multicolumn{1}{c|}{\cellcolor[HTML]{FFFFFF}\begin{tabular}[c]{@{}c@{}}0.7615\\ ($\pm$ 2.532)\end{tabular}} &
 \begin{tabular}[c]{@{}c@{}}1.0102\\ ($\pm$ 2.617)\end{tabular} &
 \multicolumn{1}{c|}{\cellcolor[HTML]{FFFFFF}\begin{tabular}[c]{@{}c@{}}0.7798\\ ($\pm$ 2.299)\end{tabular}} &
 \begin{tabular}[c]{@{}c@{}}0.9360\\ ($\pm$ 2.038)\end{tabular} &
 \cellcolor[HTML]{EFEFEF}\textbf{ToD 1, ES 6} &
 \multicolumn{1}{c|}{\cellcolor[HTML]{FFFFFF}\begin{tabular}[c]{@{}c@{}}0.6173\\ ($\pm$ 2.079)\end{tabular}} &
 \begin{tabular}[c]{@{}c@{}}0.7253\\ ($\pm$ 1.916)\end{tabular} &
 \multicolumn{1}{c|}{\cellcolor[HTML]{FFFFFF}\begin{tabular}[c]{@{}c@{}}0.7348\\ ($\pm$ 2.357)\end{tabular}} &
 \begin{tabular}[c]{@{}c@{}}0.82257\\ ($\pm$ 2.042)\end{tabular} \\ \hline
\rowcolor[HTML]{FFFFFF} 
\cellcolor[HTML]{EFEFEF}\textbf{ToD 2, ES 2} &
 \multicolumn{1}{c|}{\cellcolor[HTML]{FFFFFF}\begin{tabular}[c]{@{}c@{}}0.5833\\ ($\pm$ 2.018)\end{tabular}} &
 \begin{tabular}[c]{@{}c@{}}0.7664\\ ($\pm$ 2.116)\end{tabular} &
 \multicolumn{1}{c|}{\cellcolor[HTML]{FFFFFF}\begin{tabular}[c]{@{}c@{}}0.6039\\ ($\pm$ 1.737)\end{tabular}} &
 \begin{tabular}[c]{@{}c@{}}0.7210\\ ($\pm$ 1.37)\end{tabular} &
 \cellcolor[HTML]{EFEFEF}\textbf{ToD 2, ES 7} &
 \multicolumn{1}{c|}{\cellcolor[HTML]{FFFFFF}\begin{tabular}[c]{@{}c@{}}0.524\\ ($\pm$ 1.484)\end{tabular}} &
 \begin{tabular}[c]{@{}c@{}}0.6334\\ ($\pm$ 1.311)\end{tabular} &
 \multicolumn{1}{c|}{\cellcolor[HTML]{FFFFFF}\begin{tabular}[c]{@{}c@{}}0.5892\\ ($\pm$ 1.715)\end{tabular}} &
 \begin{tabular}[c]{@{}c@{}}0.7089\\ ($\pm$ 1.456)\end{tabular} \\ \hline
\rowcolor[HTML]{FFFFFF} 
\cellcolor[HTML]{EFEFEF}\textbf{ToD 3, ES 3} &
 \multicolumn{1}{c|}{\cellcolor[HTML]{FFFFFF}\begin{tabular}[c]{@{}c@{}}0.3968\\ ($\pm$ 1.086)\end{tabular}} &
 \begin{tabular}[c]{@{}c@{}}0.4425\\ ($\pm$ 0.745)\end{tabular} &
 \multicolumn{1}{c|}{\cellcolor[HTML]{FFFFFF}\begin{tabular}[c]{@{}c@{}}0.4284\\ ($\pm$ 1.099)\end{tabular}} &
 \begin{tabular}[c]{@{}c@{}}0.4398\\ ($\pm$ 0.538)\end{tabular} &
 \cellcolor[HTML]{EFEFEF}\textbf{ToD 3, ES 8} &
 \multicolumn{1}{c|}{\cellcolor[HTML]{FFFFFF}\begin{tabular}[c]{@{}c@{}}0.5665\\ ($\pm$ 1.801)\end{tabular}} &
 \begin{tabular}[c]{@{}c@{}}0.7323\\ ($\pm$ 1.84)\end{tabular} &
 \multicolumn{1}{c|}{\cellcolor[HTML]{FFFFFF}\begin{tabular}[c]{@{}c@{}}0.6644\\ ($\pm$ 2.214)\end{tabular}} &
 \begin{tabular}[c]{@{}c@{}}0.7498\\ ($\pm$ 1.668)\end{tabular} \\ \hline
\rowcolor[HTML]{FFFFFF} 
\cellcolor[HTML]{EFEFEF}\textbf{ToD 4, ES 4} &
 \multicolumn{1}{c|}{\cellcolor[HTML]{FFFFFF}\begin{tabular}[c]{@{}c@{}}0.5187\\ ($\pm$ 1.245)\end{tabular}} &
 \begin{tabular}[c]{@{}c@{}}0.6093\\ ($\pm$ 0.917)\end{tabular} &
 \multicolumn{1}{c|}{\cellcolor[HTML]{FFFFFF}\begin{tabular}[c]{@{}c@{}}0.6496\\ ($\pm$ 1.692)\end{tabular}} &
 \begin{tabular}[c]{@{}c@{}}07679\\ ($\pm$ 1.18)\end{tabular} &
 \cellcolor[HTML]{EFEFEF}\textbf{ToD 4, ES 9} &
 \multicolumn{1}{c|}{\cellcolor[HTML]{FFFFFF}\begin{tabular}[c]{@{}c@{}}0.5396\\ ($\pm$ 1.622)\end{tabular}} &
 \begin{tabular}[c]{@{}c@{}}0.6837\\ ($\pm$ 1.521)\end{tabular} &
 \multicolumn{1}{c|}{\cellcolor[HTML]{FFFFFF}\begin{tabular}[c]{@{}c@{}}0.578\\ ($\pm$ 1.802)\end{tabular}} &
 \begin{tabular}[c]{@{}c@{}}0.64\\ ($\pm$ 1.344)\end{tabular} \\ \hline
\rowcolor[HTML]{FFFFFF} 
\cellcolor[HTML]{EFEFEF}\textbf{ToD 5, ES 5} &
 \multicolumn{1}{c|}{\cellcolor[HTML]{FFFFFF}\begin{tabular}[c]{@{}c@{}}0.3183\\ ($\pm$ 0.78)\end{tabular}} &
 \begin{tabular}[c]{@{}c@{}}0.3497\\ ($\pm$ 0.363)\end{tabular} &
 \multicolumn{1}{c|}{\cellcolor[HTML]{FFFFFF}\begin{tabular}[c]{@{}c@{}}0.3989\\ ($\pm$ 1.051)\end{tabular}} &
 \begin{tabular}[c]{@{}c@{}}0.4387\\ ($\pm$ 0.569)\end{tabular} &
 \cellcolor[HTML]{EFEFEF}\textbf{ToD 5, ES 10} &
 \multicolumn{1}{c|}{\cellcolor[HTML]{FFFFFF}\begin{tabular}[c]{@{}c@{}}0.5441\\ ($\pm$ 1.262)\end{tabular}} &
 \begin{tabular}[c]{@{}c@{}}0.6383\\ ($\pm$ 0.791)\end{tabular} &
 \multicolumn{1}{c|}{\cellcolor[HTML]{FFFFFF}\begin{tabular}[c]{@{}c@{}}0.5576\\ ($\pm$ 1.365)\end{tabular}} &
 \begin{tabular}[c]{@{}c@{}}0.6569\\ ($\pm$ 0.925)\end{tabular} \\ \hline
\end{tabular}%
}
\caption{DISTINQT vs Centralized QoS prediction performance for 10 indicative and challenging evaluation scenarios}
\label{tab:centr_vs_distr_specific_scenarios}
\vspace{-10pt}
\end{table*}
Fig. \ref{act_pred_performance} presents the comparison between the groundtruth (actual values of uplink throughput) and the predicted output, for both QoS prediction approaches and for each of the 5 ToD-UEs. For presentation simplicity we have selected one indicative ES for each input feature configuration and for each ToD-UE. The timeline of each ES illustrates the predicted QoS value and corresponding groundtruth at intervals of 1 second ($t=1$ sec), specifically focusing on the 20th-second prediction timestep within the prediction horizon. From Fig. \ref{act_pred_performance} it is evident that both approaches are able to capture the vast majority of the uplink throughput fluctuations for the different ToD-UEs in all evaluation scenarios. 

Complementary to Fig. \ref{act_pred_performance}, Table \ref{tab:centr_vs_distr_specific_scenarios} and Fig. \ref{distr_centr_c1_vs_c2} indicate once again a statistically similar performance between our DISTINQT framework and the centralized approach in all evaluation scenarios. More specifically, Fig. \ref{distr_centr_c1_vs_c2} illustrates the performance of both approaches over different prediction horizons (in seconds). Intuitively, the further in time a prediction is made, the higher is the introduced MAE score and -as a result- the lower the model's performance. DISTINQT in almost all evaluation scenarios and prediction horizons achieves a slightly better performance under both configurations \textbf{c1} and \textbf{c2}, as it can also be inferred from the overall MAE score of Table \ref{tab:centr_vs_distr_specific_scenarios}.

In summary, our proposed DISTINQT framework can achieve and even surpass the performance of a high performing state-of-the-art LSTM-based centralized approach, considering different NETs, input feature configurations and prediction horizons. Evaluation results also showcase the contribution of the additional input features of the MEC Server NET to the overall performance of the DISTINQT.
\vspace{-5pt}
\subsection{Baseline Comparisons}
Apart from the comparison analysis between our DISTINQT framework and the centralized LSTM-based approach \cite{BarMagKour2021}, this last round of evaluation results includes a comparison with six additional state-of-the-art centralized baseline solutions for different prediction time horizons and for all evaluation scenarios of the ToD-UEs. Specifically, these baseline solutions are the following: 
\begin{itemize}
 \item \textbf{ARIMA}: An Autoregressive integrated moving average (ARIMA) model, \cite{Madan8530608}, estimating the uplink throughput solely based on the historical values of the uplink throughput. 
 \item \textbf{ARIMA-RF}: A Random Forest (RF) Regressor, estimating the uplink throughput at a specific prediction timestep, exploiting ARIMA-based input estimates \cite{9566486Kousaridas}.
 \item \textbf{RF Regressor}: A standalone RF Regressor, estimating the uplink throughput over a prediction horizon, \cite{9605036Palaios}.
 \item \textbf{k-NN Regressor, MLP \& XGBoost}: 3 different baseline solutions, namely a k-Nearest Neighbor (k-NN) Regressor, a Multi-Layer Perceptron (MLP) and an eXtreme Gradient Boosting (XGBoost), employed in the work \cite{BARMPOUNAKIS2022109341} for QoS prediction. 
\end{itemize}

\begin{figure}[h!]
 \centering
 \includegraphics[width=0.9\linewidth]{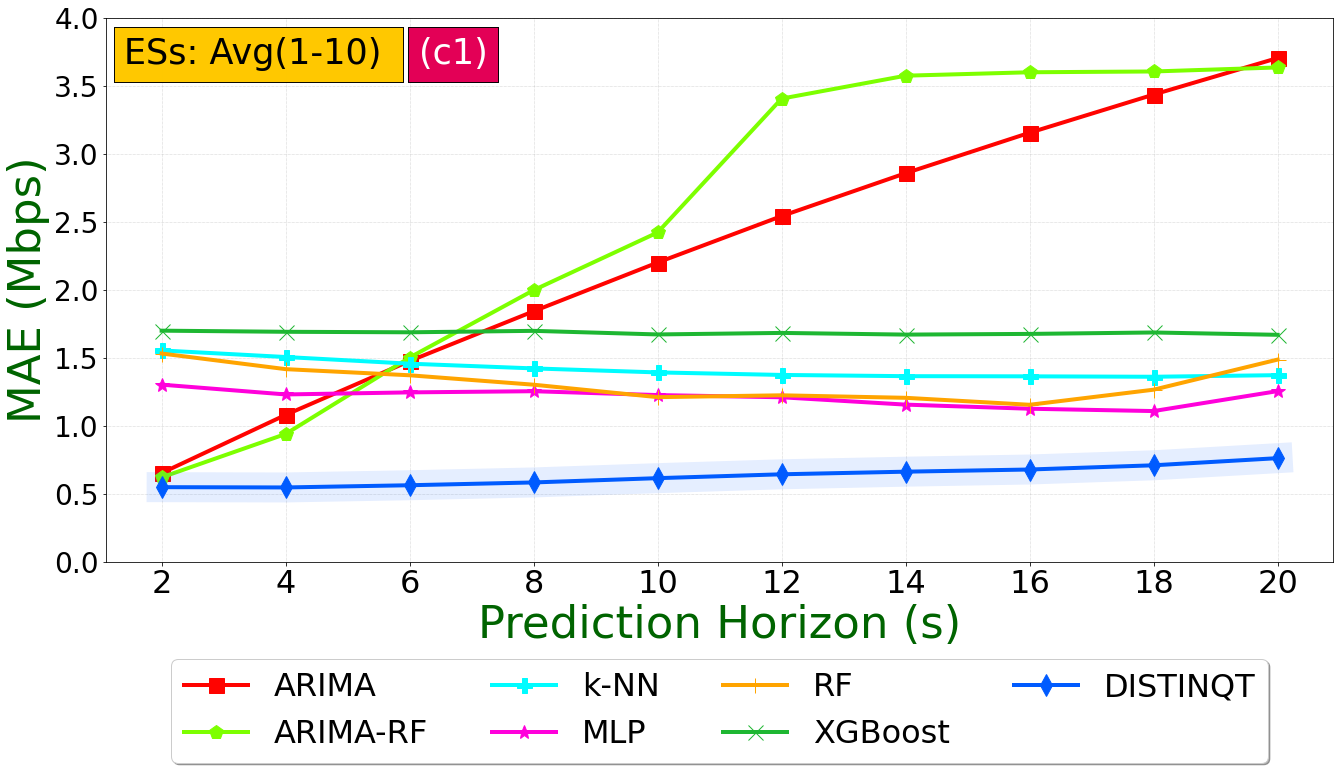}
 \caption{QoS prediction performance between DISTINQT and six centralized state-of-the-art methods for \textbf{c1}}
 \label{mae_all_baselines_4a}
 \vspace{-8pt}
\end{figure}

\begin{figure}[h!]
 \centering
 \includegraphics[width=0.9\linewidth]{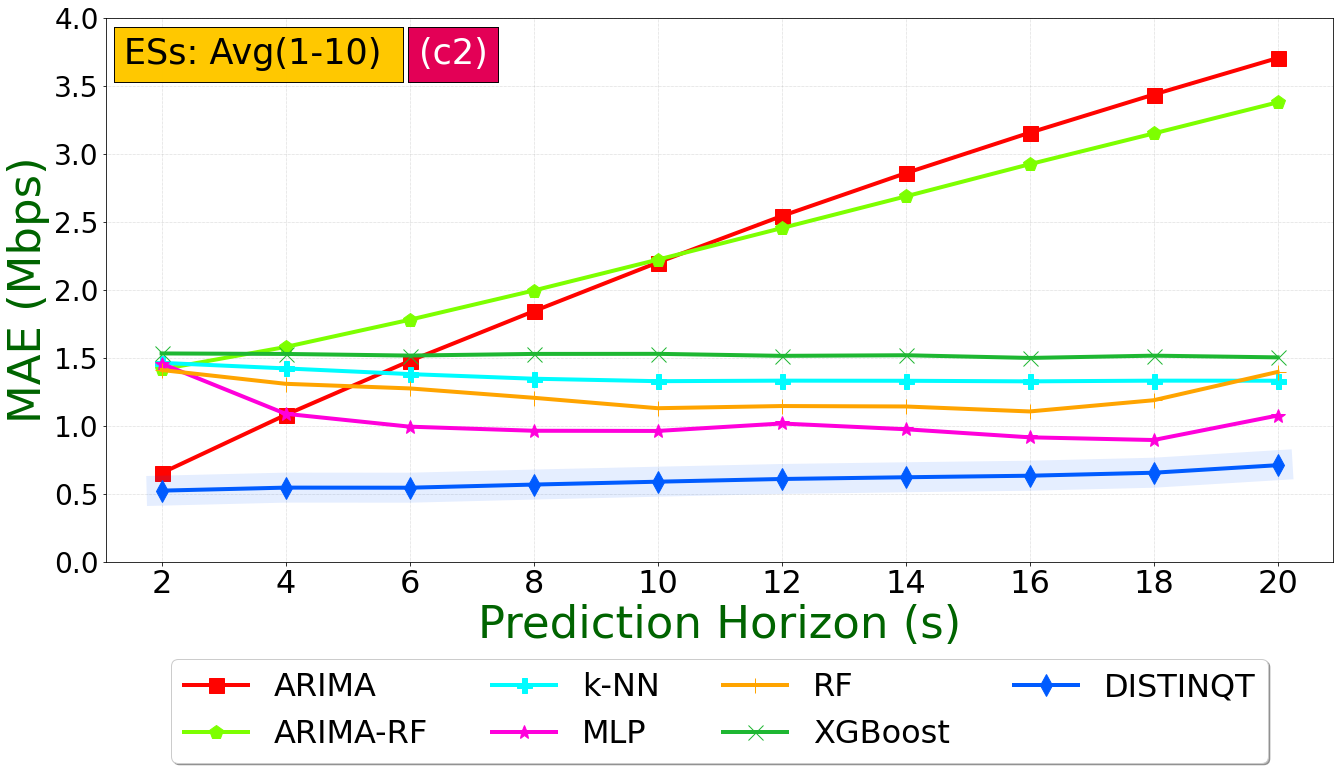}
 \caption{Qos prediction performance between DISTINQT and six centralized state-of-the-art methods for \textbf{c2}}
 \label{mae_all_baselines_4b}
 \vspace{-5pt}
\end{figure}

Figs. \ref{mae_all_baselines_4a}, \ref{mae_all_baselines_4b} and Table \ref{tab:distr_vs_baselines_all_scenarios_c1_2} present the baseline comparison results. DISTINQT outperforms all baseline solutions in all selected prediction time horizons and under both input feature configurations, retaining a stable performance as the prediction time horizon increases. Table \ref{tab:distr_vs_baselines_all_scenarios_c1_2} includes a detailed overview concerning the performance of all solutions for both input feature configurations. The table showcases the effectiveness of our approach by achieving an average performance improvement equal to 65\% and 64\% for the overall MAE score and 61.5\% and 61.1\% for the 20th second MAE under configurations \textbf{c1} and \textbf{c2}, respectively. 

\begin{table}[h!]
\renewcommand{\arraystretch}{1.05}
\centering
\resizebox{220pt}{!}{%
\begin{tabular}{|
>{\columncolor[HTML]{EFEFEF}}l |clc|cllc|}
\hline
\multicolumn{1}{|c|}{\cellcolor[HTML]{EFEFEF}} &
 \multicolumn{3}{c|}{\cellcolor[HTML]{EFEFEF}\textbf{Configuration c1}} &
 \multicolumn{4}{c|}{\cellcolor[HTML]{EFEFEF}\textbf{Configuration c2}} \\ \cline{2-8} 
\multicolumn{1}{|c|}{\multirow{-2}{*}{\cellcolor[HTML]{EFEFEF}\textbf{\begin{tabular}[c]{@{}c@{}}Avg. (Mbps)\\ ($\pm$ STD)\end{tabular}}}} &
 \multicolumn{2}{c|}{\cellcolor[HTML]{EFEFEF}\textbf{\begin{tabular}[c]{@{}c@{}}Overall\\ MAE\end{tabular}}} &
 \cellcolor[HTML]{EFEFEF}\textbf{\begin{tabular}[c]{@{}c@{}}20th Second\\ MAE\end{tabular}} &
 \multicolumn{3}{c|}{\cellcolor[HTML]{EFEFEF}\textbf{\begin{tabular}[c]{@{}c@{}}Overall\\ MAE\end{tabular}}} &
 \cellcolor[HTML]{EFEFEF}\textbf{\begin{tabular}[c]{@{}c@{}}20th Second\\ MAE\end{tabular}} \\ \hline
\cellcolor[HTML]{EFEFEF}\textbf{DISTINQT} &
 \multicolumn{2}{c|}{\cellcolor[HTML]{FFFFFF}\begin{tabular}[c]{@{}c@{}}0.554\\ ($\pm$1.578)\end{tabular}} &
 \cellcolor[HTML]{FFFFFF}\begin{tabular}[c]{@{}c@{}}0.6854\\ ($\pm$1.434)\end{tabular} &
 \multicolumn{3}{c|}{\cellcolor[HTML]{FFFFFF}\begin{tabular}[c]{@{}c@{}}0.529\\ ($\pm$1.574)\end{tabular}} &
 \cellcolor[HTML]{FFFFFF}\begin{tabular}[c]{@{}c@{}}0.6388\\ ($\pm$1.325)\end{tabular} \\ \hline
\textbf{ARIMA} &
 \multicolumn{2}{c|}{\cellcolor[HTML]{FFFFFF}\begin{tabular}[c]{@{}c@{}}2.003\\ ($\pm$4.21)\end{tabular}} &
 \cellcolor[HTML]{FFFFFF}\begin{tabular}[c]{@{}c@{}}3.4374\\ ($\pm$3.432)\end{tabular} &
 \multicolumn{3}{c|}{\cellcolor[HTML]{FFFFFF}\begin{tabular}[c]{@{}c@{}}2.003\\ ($\pm$4.21)\end{tabular}} &
 \cellcolor[HTML]{FFFFFF}\begin{tabular}[c]{@{}c@{}}3.4374\\ ($\pm$3.432)\end{tabular} \\ \hline
\textbf{ARIMA-RF} &
 \multicolumn{2}{c|}{\begin{tabular}[c]{@{}c@{}}2.3915\\ ($\pm$4.557)\end{tabular}} &
 \begin{tabular}[c]{@{}c@{}}3.6365\\ ($\pm$3.526)\end{tabular} &
 \multicolumn{3}{c|}{\begin{tabular}[c]{@{}c@{}}2.2613\\ ($\pm$4.006)\end{tabular}} &
 \begin{tabular}[c]{@{}c@{}}3.3799\\ ($\pm$3.66)\end{tabular} \\ \hline
\textbf{kNN} &
 \multicolumn{2}{c|}{\begin{tabular}[c]{@{}c@{}}1.4297\\ ($\pm$3.746)\end{tabular}} &
 \begin{tabular}[c]{@{}c@{}}1.3723\\ ($\pm$2.489)\end{tabular} &
 \multicolumn{3}{c|}{\begin{tabular}[c]{@{}c@{}}1.3665\\ ($\pm$3.413)\end{tabular}} &
 \begin{tabular}[c]{@{}c@{}}1.3315\\ ($\pm$2.79)\end{tabular} \\ \hline
\textbf{RF} &
 \multicolumn{2}{c|}{\begin{tabular}[c]{@{}c@{}}1.3249\\ ($\pm$2.791)\end{tabular}} &
 \begin{tabular}[c]{@{}c@{}}1.4869\\ ($\pm$1.539)\end{tabular} &
 \multicolumn{3}{c|}{\begin{tabular}[c]{@{}c@{}}1.2371\\ ($\pm$2.653)\end{tabular}} &
 \begin{tabular}[c]{@{}c@{}}1.396\\ ($\pm$1.532)\end{tabular} \\ \hline
\textbf{XGBoost} &
 \multicolumn{2}{c|}{\begin{tabular}[c]{@{}c@{}}1.6846\\ ($\pm$1.904)\end{tabular}} &
 \begin{tabular}[c]{@{}c@{}}1.6692\\ ($\pm$1.397)\end{tabular} &
 \multicolumn{3}{c|}{\begin{tabular}[c]{@{}c@{}}1.5202\\ ($\pm$1.76)\end{tabular}} &
 \begin{tabular}[c]{@{}c@{}}1.5032\\ ($\pm$1.247)\end{tabular} \\ \hline
\textbf{MLP} &
 \multicolumn{2}{c|}{\begin{tabular}[c]{@{}c@{}}1.2206\\ ($\pm$2.57)\end{tabular}} &
 \begin{tabular}[c]{@{}c@{}}1.255\\ ($\pm$1.231)\end{tabular} &
 \multicolumn{3}{c|}{\begin{tabular}[c]{@{}c@{}}1.0638\\ ($\pm$2.466)\end{tabular}} &
 \begin{tabular}[c]{@{}c@{}}1.0753\\ ($\pm$1.257)\end{tabular} \\ \hline
\end{tabular}%
}
\caption{QoS prediction performance comparison between DISTINQT and six baselines for \textbf{c1} \& \textbf{c2}}
\label{tab:distr_vs_baselines_all_scenarios_c1_2}
\vspace{-10pt}
\end{table}

Concluding, this section highlights the effectiveness and validity of our DISTINQT framework by showcasing that a distributed approach for QoS prediction can successfully outperform six state-of-the-art centralized solutions. 
\subsection{Data Privacy Preservation} \label{privacy_evaluation}
To evaluate the effectiveness of the proposed data privacy methods discussed in Section \ref{priv_section}, an optimization problem is examined with the objective of approximating the initial raw input data. We consider the simplest network environment, comprised of one ToD-UE and one BS worker ($e \in$ \{ToD-UE, BS\}, $k=1$) using the input features of \textbf{c1}. A privacy violator attempts to approximate the raw input data sequences $\textbf{x}_t^{e,1}$, $\forall e$, assuming: \textbf{1)} access to both ToD-UE and BS NN-Encoders, \textbf{2)} knowledge of the exchanged context vectors, and the shape of the initial raw input data sequences, and \textbf{3)} knowledge of the data normalization scale. 

The privacy violator, using a uniform distribution, initializes the estimations $\hat{\textbf{x}}_t^{e,1}$, $\forall e$, in the range 
$[0,1]$ and uses these estimations along with the respective NN-Encoders to produce the context vectors $\hat{\textbf{v}_{t}}^{(e,1)}$. The objective is to minimize the MSE $d_t^{e,1}$ between the $\hat{\textbf{v}_{t}}^{(e,1)}$ and the original ones ($\textbf{v}_{t}^{(e,1)}$), defined as:
\vspace{-7pt}
\begin{equation}
    \min_{\hat{\textbf{x}}_t^{e,1}} d_t^{e,1} = ||\textbf{v}_{t}^{(e,1)} - \hat{\textbf{v}_{t}}^{(e,1)}||^2, \forall e.
    \vspace{-5pt}
\end{equation}
To achieve this, an iterative process takes place, where the $\hat{\textbf{x}}_t^{e,1}$ ($\forall e$) is updated at each round, using the gradients $\nabla_{\hat{\textbf{x}}_t^{e,1}}{d_t^{e,1}}$ and the Adam optimizer with $lr = \textit{3e-04}$. This process is repeated until convergence, i.e. the $d_t^{e,1}$ does not improve further. To gain more insights on the similarity between the $\hat{\textbf{x}}_t^{e,1}$ and $\textbf{x}_t^{e,1}$, we use the following Euclidean Distance percentage:
\begin{equation}
    S_e = {\frac{1}{1 + u_e}}\cdot 100 \%, \forall e,
    \vspace{-5pt}
\end{equation}
where
\vspace{-5pt}
\begin{equation}
    u_e = ||\textbf{x}_t^{e,1} - \hat{\textbf{x}}_t^{e,1}||_2.
\end{equation}

\begin{figure}[h!]
    \centering
   \includegraphics[width=0.9\linewidth]{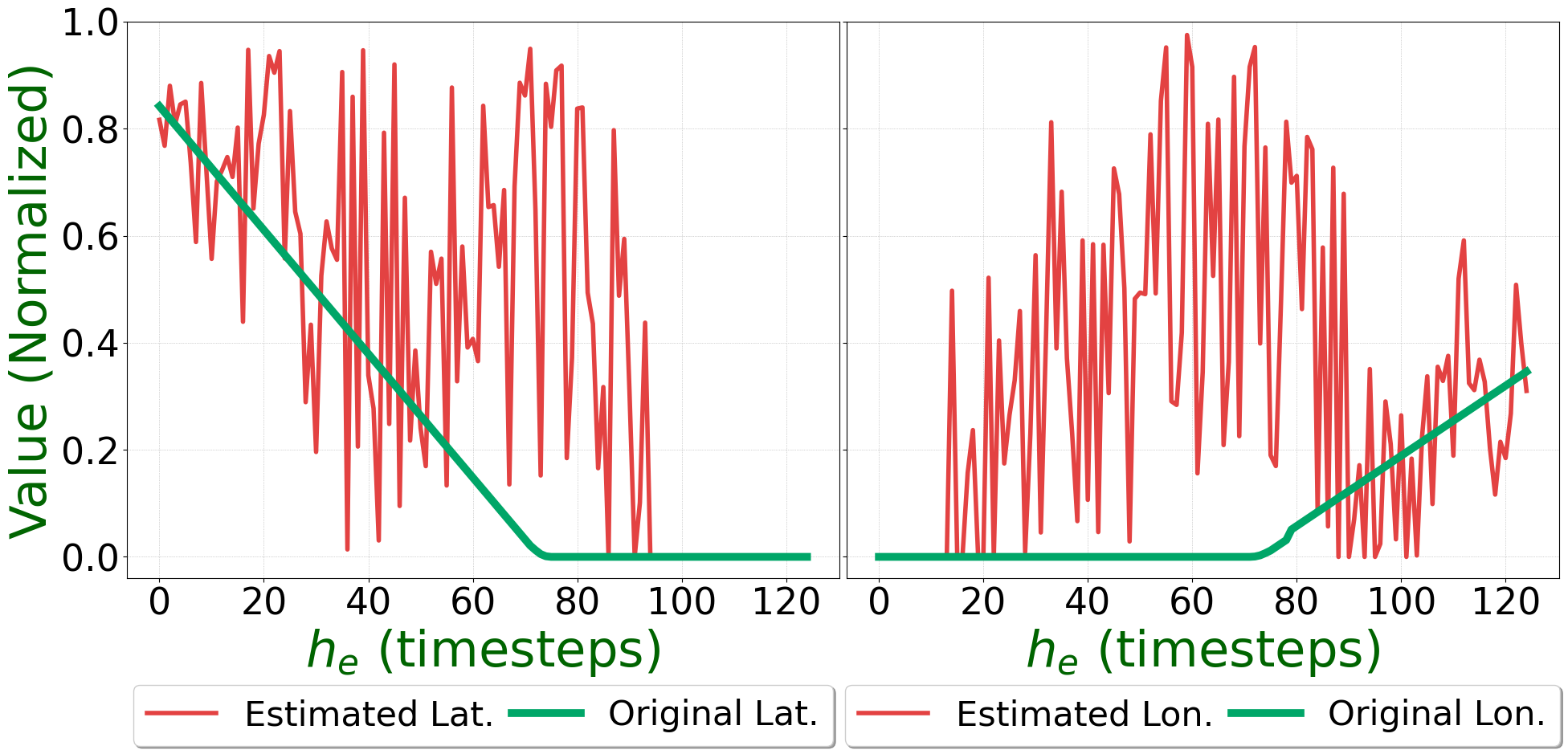}
    \caption{Original vs Estimated Sensitive Input Sequences}
    \label{privacy_exp}
    \vspace{-8pt}
\end{figure}

An experiment was conducted, where given a randomly selected historical window for the $\textbf{x}_t^{e,1}, \forall e$, the privacy violator after $200$K iterations converged to a $d_t^{\text{ToD-UE},1} = \textit{3.98e-09}$ and $d_t^{\text{BS},1} = \textit{3.82e-10}$ for the ToD-UE and the BS, respectively. 
Fig. \ref{privacy_exp} depicts a comparison between the original and estimated sensitive data (i.e. the location history of the ToD-UE), indicating a low similarity percentage equal to $20.26\%$ and $17.84\%$ for the latitude and longitude, respectively.
The overall similarity percentages for all features and for both NETs are approximately equal to $S_{\text{ToD-UE}} = 18.69\%$ and $S_{\text{BS}}=23.17\%$. 

It is important to note that, based on the results presented above, the estimation of the normalized input sequence is a challenging task. The lack of knowledge regarding the normalization method employed and the value range of each feature in each NET makes it impossible to obtain an initial raw input sequence that can be considered accurate.

\section{Conclusions} \label{conclusions}
This paper introduces and demonstrate the effectiveness of DISTINQT, a novel privacy-aware distributed learning framework for QoS prediction in the presence of heterogeneous data types and model architectures. DISTINQT is based on sequence-to-sequence autoencoders, distributing various NN-Encoders across the network and contributing to data privacy preservation. Our framework is evaluated in the Tele-Operated Driving use case, where evaluation results show that it can reach and even surpass the performance of its centralized counterpart, while proving the validity of the privacy preserving claims. This is especially important for such safety critical use cases that require immediate and accurate decision making without compromising any sensitive information. Finally, DISTINQT achieves an average decrease in prediction error up to 65\%, compared to six state-of-the-art centralized baseline solutions. Our future work will focus on finding a balance between the communication overhead that is introduced in the learning process and the robustness and generalization capabilities of the model.
\vspace{-4pt}
\bibliographystyle{IEEEtran}
\bibliography{bibliography/bib}

\end{document}